%% file: ftqc-cs-main.tex
\documentclass[aps,pra,floatfix,showpacs,twocolumn,nofootinbib,superscriptaddress]{revtex4}

\usepackage{epsfig}
\usepackage{amsmath}

\newtheorem{theorem}{Theorem}
\newtheorem{proposition}{Proposition}
\newtheorem{lemma}{Lemma}

\begin{document}

\title{Fault-tolerant quantum computation with cluster states}

\author{Michael~A.~Nielsen}
\email{nielsen@physics.uq.edu.au}
\homepage{www.qinfo.org/people/nielsen/}
\affiliation{School of Physical Sciences, The University of
  Queensland, Brisbane, Queensland 4072, Australia.}
\affiliation{School of Information Technology and Electrical
  Engineering, The University of Queensland, Brisbane, Queensland
  4072, Australia.}

\author{Christopher~M.~Dawson}
\email{dawson@physics.uq.edu.au}
\homepage{www.physics.uq.edu.au/people/dawson/}
\affiliation{School of Physical Sciences, The University of
  Queensland, Brisbane, Queensland 4072, Australia.}

\date{\today}


\begin{abstract}
  The one-way quantum computing model introduced by Raussendorf and
  Briegel [Phys. Rev. Lett. \textbf{86} (22), 5188-5191 (2001)] shows
  that it is possible to quantum compute using only a fixed entangled
  resource known as a \emph{cluster state}, and adaptive single-qubit
  measurements.  This model is the basis for several practical
  proposals for quantum computation, including a promising proposal
  for optical quantum computation based on cluster states [M.  A.
  Nielsen, arXiv:quant-ph/0402005, accepted to appear in
  Phys.~Rev.~Lett.].  A significant open question is whether such
  proposals are scalable in the presence of physically realistic
  noise.  In this paper we prove two threshold theorems which show
  that scalable fault-tolerant quantum computation may be achieved in
  implementations based on cluster states, provided the noise in the
  implementations is below some constant \emph{threshold} value.  Our
  first threshold theorem applies to a class of implementations in
  which entangling gates are applied \emph{deterministically}, but
  with a small amount of noise.  We expect this threshold to be
  applicable in a wide variety of physical systems.  Our second
  threshold theorem is specifically adapted to proposals such as the
  optical cluster-state proposal, in which \emph{non-deterministic}
  entangling gates are used.  A critical technical component of our
  proofs is two powerful theorems which relate the properties of noisy
  unitary operations restricted to act on a subspace of state space to
  extensions of those operations acting on the entire state space.  We
  expect these theorems to have a variety of applications in other
  areas of quantum information science.
\end{abstract}

\maketitle

\input{ftqc-cs-intro}
\input{ftqc-cs-noise}
\input{ftqc-cs-csc}
\input{ftqc-cs-deterministic}

\input{ftqc-cs-ocs}

\input{ftqc-cs-conclusion}

\appendix

\input{ftqc-cs-app1}

\input{ftqc-cs-app2}

\newpage
%=====================================================================
%\bibliographystyle{prsty}
\bibliography{./mybib}

\end{document}

%% file: ftqc-cs-intro.tex
\section{Introduction}

\subsection{Overview}

%
% into paragraph
%
One of the most surprising recent developments in quantum computation
is the insight that \emph{quantum measurement} can be used as the
fundamental dynamical operation in a quantum
computer~\cite{Raussendorf01a,Nielsen03b}.  This insight has
significant implications for our theoretical understanding of how
quantum computers operate, and also for the development of practical
proposals for quantum computing.

%
% overview of the history: cluster-state model
%
Historically, the first measurement-based model for quantum computing
was the \emph{one-way quantum computer} developed by Raussendorf and
Briegel~\cite{Raussendorf01a}.  A one-way quantum computation is
performed in two stages.  In the first stage an entangled many-qubit
state known as the \emph{cluster state} is prepared.  This is a fixed
entangled state that does not depend on the problem instance being
solved by the computation.  Indeed, when the one-way quantum
computation is being used to simulate a quantum circuit\footnote{For a
  review of the quantum circuit model of quantum computation,
  see~\cite{Nielsen00a}.},~\cite{Raussendorf01a} shows that the
identity of the cluster state can be made very nearly independent of
the details of the circuit being simulated, with the only dependence
being on the \emph{depth} and \emph{breadth} of the circuit.  In the
second stage of a one-way quantum computation a sequence of
single-qubit measurements is performed on the cluster state.  These
measurements are \emph{adaptive}, in the sense that the basis in which
a qubit is measured may depend upon the outcome of earlier
measurements.  Remarkably, this two-stage process is sufficient to
simulate any quantum circuit whatsoever.

%
% MAN's approach
%
More recently, an apparently quite different
teleportation-based~\cite{Bennett93a} approach to measurement-based
quantum computation was developed by Nielsen~\cite{Nielsen03b}.  This
approach is based on the idea now known as gate teleportation,
introduced by Nielsen and Chuang~\cite{Nielsen97c}, and further
developed by Gottesman and Chuang~\cite{Gottesman99a}.  When it was
first introduced, teleportation-based quantum computation appeared to
be quite different from the one-way quantum computer, but subsequent
work~\cite{Aliferis04a,Jorrand04a,Childs04a} has provided a unified
conceptual framework in which both approaches may be understood.

%
% bridge to noise
%
Although the measurement-based models of quantum computation represent
an important conceptual advance, there is an important caveat, namely,
that the measurement-based models assume all operations are carried
out perfectly.  Since real physical systems suffer from noise, to make
measurement-based models scalable we must develop techniques for
combatting noise in those models.

%
% for quantum circuits
%
The challenge posed by noise has been met in the quantum circuit model
of computation with the development of an impressive theory of
\emph{fault-tolerance}, providing a large body of techniques which can
be used to reduce the effects of noise on quantum circuits.  The
culmination of the theory of fault-tolerance is the \emph{threshold
  theorem}, which states that for physically reasonable models of
noise, and provided the noise is below some constant \emph{threshold}
value, it is possible to use quantum circuits to efficiently simulate
an arbitrarily long quantum computation with arbitrary accuracy.
Although the threshold remains to be experimentally confirmed, there
is now general agreement that, based on our best theoretical
understanding of quantum noise, the threshold theorem solves the
problem of noise in the quantum circuit model of computation.  That
is, quantum noise poses no problem of principle for quantum
computation, only the (very significant) practical problem of reducing
noise levels below the threshold value.  For a survey of the theory of
fault-tolerance and the threshold theorem see Chapter~10
of~\cite{Nielsen00a}, and references therein.

%
% noise in cluster states
%
The purpose of the present paper is to develop similar fault-tolerant
threshold results for several measurement-based models of quantum
computation, focusing primarily on models derived from the one-way
quantum computer of Raussendorf and Briegel.  We refer to this entire
class of models as the \emph{cluster-state model of quantum
  computation}, to emphasize the crucial role played by cluster
states.  Note that we use the term ``cluster-state model of quantum
computation'' in a rather loose sense, using it to denote an entire
class of models based on cluster states.  We reserve the term
\emph{one-way quantum computer} to refer to the specific model
originally suggested in~\cite{Raussendorf01a}.

%
% our motivation
%
Our primary motivation in studying fault-tolerance in the
cluster-state model is to establish that the cluster-state model can
be used as the basis for \emph{scalable} practical proposals for
quantum computation.  Although several proposals for experimental
quantum computation with cluster states have been
made~\cite{Raussendorf01a,Nielsen04b}, such proposals cannot be
considered scalable unless fault-tolerant methods of implementation
are developed.  In this paper we develop methods for fault-tolerant
computation using cluster states, methods that are applicable in a
wide variety of practical proposals.

%
% prior work by R&B
%
Prior work on the problem of fault-tolerant computation with cluster
states has been reported in Chapter~4 of Raussendorf's
thesis~\cite{Raussendorf03b}.  This work obtained a threshold for a
class of noise models in which Pauli errors occur probabilistically in
a cluster-state computation.  Our threshold result applies to a more
general noise model that is likely to be more realistic in many
physical systems; subject to some assumptions about locality, we allow
arbitrary non-Markovian noise to occur in the computation, and even
allow errors to occur in the accompanying classical computation.
Thus, our work should be viewed as extending and complementing the
approach taken in~\cite{Raussendorf03b}.  We note that independent
work extending~\cite{Raussendorf03b} is also being undertaken by
Raussendorf and Briegel~\cite{Raussendorf04a}.

%
% what makes things difficult I
%
What is it that makes proving a fault-tolerant threshold in the
cluster-state model non-trivial?  The obvious approach to proving a
threshold is to take the quantum circuit that we want to make
fault-tolerant, convert it to a fault-tolerant quantum circuit using
the standard prescriptions, and then simulate the resulting circuit
using a cluster-state computation.  It seems physically plausible that
noise occurring in the cluster-state model of computation should then
be corrected by the error-correcting properties of the original
fault-tolerant circuit.

%
% what makes things difficult II
%
Two difficulties obstruct this proposal.  The first difficulty is that
the qubits in the cluster tend to degrade before they are measured.
We will see that this difficulty can easily be overcome by building up
the cluster in parts, so that no part of the cluster is allowed to
degrade too much before being measured.

%
% what makes things difficult III
%
The second difficulty is more serious. While it is plausible that
noise in the cluster-state computation is corrected by the
error-correcting properties of the original fault-tolerant circuit,
showing this turns out to be non-trivial.  The key is to prove that
noise in a cluster-state simulation of a quantum circuit can be mapped
onto \emph{equivalent noise} in the quantum circuit.  Provided that
mapping has suitable properties, a threshold theorem then follows.
The greater part of this paper is spent in constructing such a
mapping.  To underscore the difficulty in proving this, we mention
just one interesting subtlety: we shall see that Markovian noise in a
cluster-state computation actually maps onto non-Markovian noise in
the corresponding quantum circuit.  This and other subtleties make the
task of proving a threshold technically challenging.

%
% the fact that we don't prove a threshold
%
This discussion highlights a general point worth noting.  Our paper
provides a way of taking an arbitrary quantum circuit and then
simulating it in a fault-tolerant fashion in the cluster-state model
of computation.  We don't, however, provide a direct way of making a
\emph{cluster-state} computation fault-tolerant, except insofar as a
cluster-state computation may be regarded as a special type of quantum
circuit computation.  It would be interesting to investigate more
direct fault-tolerance constructions applicable to an arbitrary
cluster-state computation.  Of course, for the purposes of proving the
feasibility of scalable quantum computation in the cluster-state
model, the present approach is sufficient.

\subsection{Optical cluster states and fault tolerance}

%
% introduce optical cluster
%
A topic of special interest in the present paper is the optical
cluster-state proposal for quantum computation suggested by
Nielsen~\cite{Nielsen04b}.  Optical systems offer a number of
significant experimental advantages for quantum computing, and this
proposal thus offers a very promising approach to experimental quantum
computation.  However, the optical cluster-state proposal also differs
from most quantum computing proposals in that it is based on
entangling gates that only work \emph{non-deterministically}.  This
non-deterministic nature poses special difficulties when attempting to
prove a threshold for the optical cluster-state proposal.
Following~\cite{Nielsen04b}, we now briefly review some background on
this proposal that will help the reader understand how it fits into
the present paper.

%
% basic optics
%
\emph{A priori}, optics offers significant advantages for the
implementation of quantum computation, such as the ease of performing
basic manipulations, and long decoherence times.  Unfortunately,
standard linear optical elements alone are unsuitable for quantum
computation, as they do not enable photons to interact.  This
difficulty can, in principle, be resolved by making use of nonlinear
optical elements~\cite{Yamamoto88a,Milburn89a}, at the price of
requiring large nonlinearities that are at present extremely difficult
to achieve.

%
% KLM
%
An alternate approach was developed by Knill, Laflamme and Milburn
(KLM)~\cite{Knill01a}, who proposed using measurement to effect
entangling interactions between optical qubits.  Using this idea, KLM
developed a scheme for scalable quantum computation based on linear
optical elements, together with high-efficiency photodetection,
feedforward of measurement results, and single-photon generation.  KLM
thus showed that scalable optical quantum computation is in principle
possible using relatively modest resources. Experimental
demonstrations~\cite{Pittman03a,OBrien03a,Sanaka03a,Gasparoni04a,Zhao04a}
of several of the basic elements of KLM have already been
achieved.

%
% problems
%
Despite these impressive successes, the obstacles to fully scalable
quantum computation with KLM remain formidable.  The biggest challenge
is to perform a two-qubit entangling gate in the near-deterministic
fashion required for scalable quantum computation.  KLM propose doing
this using a combination of three ideas.  (1) Using linear optics,
single-photon sources and photodetectors, \emph{non-deterministically}
perform an entangling gate.  This gate fails most of the time,
destroying the state of the computer when it does so, and so is not
immediately suitable for quantum computation.  Several variants of
this gate have already been experimentally
demonstrated~\cite{Pittman03a,OBrien03a,Sanaka03a,Gasparoni04a,Zhao04a}.
(2) By combining the basic non-deterministic gate with quantum
teleportation, a class of non-deterministic gates which are not so
destructive of the state of the computer is found.  (3) By combining
the gates from (2) with ideas from quantum error-correction, the
probability of the gate succeeding can be improved until the gate is
near-deterministic, allowing scalable quantum computation.

%
% Nielsen proposal
%
The combination of these three ideas allows scalable quantum
computation, in principle.  In practice there are enormous obstacles
to performing even a single near-deterministic entangling quantum gate
in this fashion.  The proposal of~\cite{Nielsen04b} eliminates much of
the difficulty, completely removing step (3), and obviating the need
for all but the simplest versions of step (2)\footnote{Another
  promising proposal for optical quantum computing which shares these
  attributes is~\cite{Yoran03a}.}.  This is achieved by combining some
of the simplest elements of KLM with the cluster-state model of
quantum computation.  The resulting proposal puts near-deterministic
entangling quantum gates within experimental reach, and thus offers an
extremely promising approach to quantum computation, provided suitable
methods for dealing with noise can be developed.

\subsection{Content of the paper}

%
% the results in this paper
%
In this paper we prove two threshold theorems for cluster-state
quantum computation.  The first is a general threshold theorem
applicable to a variety of possible implementations of cluster-state
computation, assuming that deterministic entangling gates are
available.  The second is specifically adapted to the optical
cluster-state proposal for quantum computation.  Taken together, these
theorems show that for a wide variety of possible physical
implementations, noise poses no problem of principle for fully
scalable cluster-state quantum computation.

%
% caveat about the paper
%
Before describing in detail the structure of the paper, it is worth
noting two issues that we do not fully address.  The first of these
issues is the determination of a numerical value for the threshold.
Although we do obtain bounds on the threshold, those bounds are
obtained through analytic methods that are far too pessimistic.  Our
philosophy is that the problem of understanding the threshold is best
split into two parts.  In the first part, one attempts to rigorously
prove the existence of a finite threshold for some large class of
noise models.  In the second part, one attempts through a combination
of numerical and analytic work to obtain a realistic estimate of the
threshold for some specific and physically-motivated noise model.  In
this second part it is much more reasonable to rely on numerical
evidence and heuristic reasoning, since results for specific noise
models can always be checked by computer simulation (and ultimately by
experiment).  Examples of this kind of work for quantum circuits may
be found in~\cite{Knill04a,Knill04b,Steane02a,Steane03a,Dur03a}, and
references therein.  Our focus in the present paper has been on the
first part of this program, obtaining a rigorous proof that a finite
threshold exists.  Detailed numerical simulation and optimization of
the threshold value for realistic noise models is underway, and will
be reported elsewhere~\cite{Dawson04b}.

%
% second issue
%
The second issue not fully addressed in this paper relates to the
noise model used in our analysis of the optical cluster-state proposal
for quantum computation.  Physically, one of the most significant
sources of noise in any optical implementation is likely to be photon
loss.  This causes the state of the optical qubit to ``leak'' from the
degrees of freedom associated with the qubit out into some other
dimensions of the physical state space.  In the context of threshold
theorems, such noise is known as a ``leakage error'', and there are
standard techniques for dealing with such errors in the theory of
fault-tolerance.  However, our threshold analysis for the
cluster-state model is based on the recent threshold theorem proved by
Terhal and Burkard~\cite{Terhal04a}, and that threshold does not
explicitly deal with leakage errors.  While it seems extremely likely
to us that the result of~\cite{Terhal04a} can be patched so that
leakage errors are accounted for, we have not worked through the
analysis in detail.  Rather than do so, in this paper we restrict
ourselves to a brief discussion of leakage, deferring full
investigation of this issue to a future publication.  The other
alternative, of course, would be to base our analysis on an
alternative version of the threshold theorem.  However, we will see
that the theorem of Terhal and Burkard has considerable advantages for
the analysis of fault-tolerance with cluster states, as the only
published version of the threshold formulated specifically to deal
with non-Markovian noise.

%
% survey of the paper: error strength
%
The structure of the paper is as follows.  We begin in
Section~\ref{sec:noise} by defining a measure of how much noise occurs
in a quantum information processing task.  We call this measure the
\emph{error strength}, and prove several simple properties of the
error strength that will be useful later in the paper.  

%
% the unitary extension theorems
%
Section~\ref{sec:noise} also contains two important technical results,
which we dub the \emph{first} and \emph{second unitary extension
  theorems}.  Roughly speaking, these results are applicable to
situations in which two unitary operations $U$ and $V$ act in a
similar fashion on a subspace $S$ of state space.  Of course, just
because $U$ and $V$ act similarly on a subspace, it does not follow
that they have similar global actions on state space.  However, the
theorems we prove guarantee that there exist \emph{unitary} extensions
$\tilde U$ and $\tilde V$ of the restrictions $U|_S$ and $V|_S$,
respectively, such that $\tilde U$ and $\tilde V$ have approximately
equal actions everywhere on state space.  These unitary extension
theorems are critical to our later analysis of fault-tolerance.  More
generally, we believe that these results are of substantial interest
independent of their application to fault-tolerance, and likely to
find application in other areas of quantum information science.

%
% review of the threshold theorem
%
Section~\ref{sec:noise} concludes with a review of the content of the
threshold theorem for quantum circuits.  We focus our attention on the
threshold theorem proved recently by Terhal and
Burkard~\cite{Terhal04a}, extending earlier work of Aharonov and
Ben-Or~\cite{Aharonov97a,Aharonov99a}.  The threshold
of~\cite{Terhal04a} is unique in that it is specifically designed with
non-Markovian noise in mind.  While several of the other known
variants of the threshold theorem can cope with some level of
non-Markovian noise, those other variants are designed primarily with
the case of Markovian noise in mind.  This is important for us as we
will see that non-Markovian noise arises naturally in the analysis of
fault-tolerant cluster-state quantum computation, even if the actual
physical noise occurring in the cluster-state computation is
Markovian.

%
% cluster state section
%
In Section~\ref{sec:cluster-state-qc} we describe the cluster-state
model of quantum computation.  Rather than providing a detailed proof
of how the model works (which is available elsewhere) we describe the
model through some simple examples.  We also discuss ways of
alleviating one of the key difficulties that arises when attempting to
perform fault-tolerant quantum computation with cluster states, the
tendency of qubits in the cluster to degrade before they are measured.
We conclude with a brief review of how the cluster-state model of
quantum computation can be combined with the ideas of KLM to obtain a
scheme for optical quantum computation.

%
% section on deterministic model
%
Section~\ref{sec:deterministic} is the heart of the paper, stating and
proving our first threshold theorem for cluster-state computation.
More precisely, what we prove is that it is possible to simulate an
arbitrary quantum circuit using a noisy cluster-state computation,
with arbitrary accuracy and only a small overhead in the physical
resources required, subject to reasonable constraints on the physical
resources available, and on the noise afflicting the implementation.

%
% sequence of ideas
%
The sequence of ideas used to prove this threshold theorem is
conceptually rather simple.  Step one is to translate the quantum
circuit that we want to simulate into a fault-tolerant circuit, using
the standard prescriptions for making a circuit fault-tolerant.  At
this stage we assume there is no noise in the computation.  Step two
is to translate the fault-tolerant circuit into a cluster-state
computation, again using standard prescriptions.  Step three is to
carefully specify a procedure for physically implementing the
cluster-state computation, a procedure that avoids the degradation of
parts of the cluster that was mentioned above.  The idea is to build
up only part of the cluster at any given time, adding extra qubits
into the cluster as required.  Up until this step we assume that all
operations are perfect.  Step four is to introduce noise into the
description of the cluster-state computation, as would occur in an
actual implementation.  The most complex and critical step, step five,
is to show that the noisy cluster-state computation is equivalent to
the \emph{original} fault-tolerant circuit, with some noise added into
the circuit.  That is, we want to map noise in the cluster-state
computation back onto \emph{equivalent noise} in the original
fault-tolerant circuit.  We will see that this mapping has the
property that provided the noise in the cluster-state model is of an
appropriate form, and not too strong, it is \emph{equivalent} to noise
in the original fault-tolerant circuit which is only slightly
stronger, and which is of a form which can be suppressed by the usual
fault-tolerance constructions for quantum circuits.  This mapping of
noise models thus enables us to infer a threshold theorem for noisy
cluster-state computation.

%
% section on fault-tolerance in the optical cluster state model
%
Section~\ref{sec:optical-ft} extends these ideas to the optical
cluster-state proposal for quantum computation.  The reason the
results of Section~\ref{sec:deterministic} cannot be immediately
applied is that the entangling gates used in the optical cluster-state
proposal are non-deterministic.  We resolve this problem by devising
an approach to computation in which the non-deterministic entangling
gates can be treated as deterministic entangling gates, subject to a
small amount of additional noise.  This enables us to map noise in the
optical cluster-state proposal into equivalent noise in the
deterministic cluster-state model, and then use the result of
Section~\ref{sec:deterministic} to infer a threshold theorem for
optical cluster-state quantum computation.

%
% conclusion
%
Section~\ref{sec:conclusion} concludes the paper with a summary of
results, and a discussion of the outlook for further developments.

%% file: ftqc-cs-noise.tex
\section{Noise and fault-tolerant quantum circuits}
\label{sec:noise}

%
% overview
%
In order to prove a threshold for cluster-state computation we first
need a way of describing quantum noise, and quantifying its effects.
In Subsection~\ref{subsec:error-strength} we introduce a measure that
quantifies the effects of quantum noise, describe some properties of
that measure, and describe the unitary extension theorems.  In
Subsection~\ref{subsec:circuit-threshold} we review the threshold
theorem for quantum circuits proved by Terhal and
Burkard~\cite{Terhal04a}.

\subsection{Error strength}
\label{subsec:error-strength}

%
% our basic noise model
%
Suppose we have a quantum system, $Q$, and we wish to implement a
unitary operation $U_Q$ on that system.  Unfortunately, the system is
not completely isolated from its environment, $E$, and thus the true
evolution of the system will be described by some unitary evolution
$V_{QE}$ acting on both the system and the environment.  We define the
\emph{error strength} or \emph{noise strength} of this operation by:
\begin{eqnarray}
  \Delta_{Q:E}(U_Q,V_{QE}) & \equiv & \min_{U_E} \| V_{QE}- U_Q \otimes U_E \|,
\end{eqnarray}
where the minimum is over all unitary operations $U_E$ on the system
$E$, and the norm is the usual operator norm.

%
% the status of this noise measure
%
We will use the error strength $\Delta_{Q:E}$ as our principal measure
of noise in the implementation of quantum computation, whether it be
by cluster-state methods, or by quantum circuits.  Although
$\Delta_{Q:E}$ has been defined only when the ideal operation $U_Q$ is
unitary, we'll see later that it can also be used to understand noise
in operations that may not be unitary, such as measurement and state
preparation.  Our main reason for using the measure $\Delta_{Q:E}$ is
that it is the same measure that Terhal and Burkard use in their study
of fault-tolerance~\cite{Terhal04a} for non-Markovian noise models.
We now describe several properties of $\Delta_{Q:E}$ that are useful
later in the paper.

\begin{proposition} {} \label{prop:chaining}
 \textbf{(Chaining property)} Let $U_Q^1,\ldots,U_Q^m$
  be unitary operations on the system $Q$, and
  $V_{QE}^1,\ldots,V_{QE}^m$ be unitary operations on the combined
  system $QE$.  Then
  \begin{eqnarray}
    \Delta_{Q:E}(U_Q^1\ldots U_Q^m,V_{QE}^1\ldots V_{QE}^m) & \leq &
    \sum_{j=1}^m \Delta_{Q:E}(U_Q^j,V_{QE}^j). \nonumber \\
 & &
  \end{eqnarray}
\end{proposition}

The chaining property tells us that the total error strength of a
sequence of imperfect operations is less than or equal to the sum of
the individual error strengths.  We note that this proposition and its
proof is similar to Lemma~1 in Section~1.3 of~\cite{Terhal04a}.

\textbf{Proof:} Choose $U_E^j$ so that $\Delta_{Q:E}(U_Q^j,V_{QE}^j) =
\| V_{QE}^j - U_Q^j \otimes U_E^j\|$, and define $\Delta^j \equiv
V_{QE}^j-U_Q^j \otimes U_E^j$.  A straightforward induction on $m$ can
be used to establish the formula
\begin{eqnarray} \label{eq:exact-form-product}
  V_{QE}^1\ldots V_{QE}^m & = & \left( U_Q^1 \ldots U_Q^m \right) \otimes
  \left( U_E^1 \ldots U_E^m \right) \nonumber \\
  & & + \sum_{j=1}^m
  \left( U_Q^1 \ldots U_Q^{j-1} \otimes
  U_E^1 \ldots U_E^{j-1} \right) \times \nonumber \\
  & & \hphantom{+ \sum_{j=1}^m} \Delta^j \,
  V_{QE}^{j+1} \ldots V_{QE}^m.
\end{eqnarray}
The result follows by subtracting $\left( U_Q^1 \ldots U_Q^m \right)
\otimes \left( U_E^1 \ldots U_E^m \right)$ from both sides of
Eq.~(\ref{eq:exact-form-product}), and applying the triangle
inequality.

\textbf{QED}

%
% noise reduction proposition
%
Later in the paper we will show that a noisy cluster-state computation
is equivalent to a noisy quantum circuit computation.  A technique
we'll use in the proof of this fact is to change the set of systems
considered to be part of the environment.  The following two
propositions help us understand the behaviour of the error strength
when the environment is changed in this way.

\begin{proposition} {} \label{prop:error-reduction}
  Let $A, B$ and $C$ be three quantum systems, and let $U_A, U_B,
  V_{ABC}$ be unitary operations acting on systems indicated by the
  respective subscripts.  Then
  \begin{eqnarray}
    \Delta_{A:BC}(U_A,V_{ABC}) \leq \Delta_{AB:C}(U_A \otimes U_B, V_{ABC}).
  \end{eqnarray}
\end{proposition}

\textbf{Proof:} The proof is immediate from the definition of
$\Delta_{Q:E}$ and the fact that the set of unitary matrices $U_{BC}$
on $BC$ is a superset of the set of unitary matrices of the form $U_B
\otimes U_C$, where $U_B$ is the (fixed) given matrix:
\begin{eqnarray}
  \Delta_{A:BC}(U_A,V_{ABC}) & = &
  \min_{U_{BC}} \| V_{ABC} - U_A \otimes U_{BC}\| \\
  & \leq & \min_{U_C} \| V_{ABC}- U_A \otimes U_B \otimes U_C \| \nonumber \\
 & & \\
  & = & \Delta_{AB:C}(U_A \otimes U_B,V_{ABC}).
\end{eqnarray}

\textbf{QED}

\begin{proposition} {} \label{prop:error-ommission}
  Let $A, B$ and $C$ be three quantum systems, and let $U_A, V_{AB},
  V_{C}$ be unitary operations acting on systems indicated by the
  respective subscripts.  Then
  \begin{eqnarray}
    \Delta_{A:BC}(U_A,V_{AB} \otimes V_C) \leq
    \Delta_{A:B}(U_A, V_{AB}).
  \end{eqnarray}
\end{proposition}

\textbf{Proof:} Similarly to the proof of the previous proposition,
the proof is immediate from the definitions and the fact that the set
of unitary matrices $U_{BC}$ on $BC$ is a superset of the set of
unitary matrices of the form $U_B \otimes V_C$, where $V_C$ is the
(fixed) given matrix:
\begin{eqnarray}
  & & \Delta_{A:BC}(U_A,V_{AB} \otimes V_C) \nonumber \\
  & = &
  \min_{U_{BC}} \| V_{AB} \otimes V_C - U_A \otimes U_{BC}\| \\
  & \leq & \min_{U_B} \| V_{AB} \otimes V_C- U_A \otimes U_B \otimes V_C \|
  \\
  & = & \min_{U_B} \| V_{AB} - U_A \otimes U_B \| \\
  & = & \Delta_{A:B}(U_A,V_{AB}).
\end{eqnarray}

\textbf{QED}

The next proposition helps in commuting noisy operations past one
another.

\begin{proposition} {} \label{prop:commute-forward}
  Let $U_Q$ and $V_Q$ be commuting unitary operations on a quantum
  system $Q$.  Let $U_{QE}$ and $V_{QE}$ be noisy versions of these
  operations involving also an environment $E$.  Then there exist
  unitaries $\tilde U_{QE}$ and $\tilde V_{QE}$ such that (a) $\tilde
  U_{QE} \tilde V_{QE} = V_{QE} U_{QE}$; $\Delta_{Q:E}(U_Q,\tilde
  U_{QE}) \leq \Delta_{Q:E}(V_Q,V_{QE})$; and $\Delta_{Q:E}(V_Q,\tilde
  V_{QE}) \leq \Delta_{Q:E}(U_Q,U_{QE})$.
\end{proposition}

This proposition tells us that if $U_Q$ and $V_Q$ commute, then
applying a noisy version of $U_Q$ followed by a noisy version of $V_Q$
is equivalent to applying a noisy version of $V_Q$ followed by a noisy
version of $U_Q$.  Furthermore, the noise strengths in the new
versions of $U_Q$ and $V_Q$ are no worse than in the originals, except
for interchanging the role of $U_Q$ and $V_Q$.

\textbf{Proof:} Using the definition of $\Delta_{Q:E}$ we may choose
unitaries $U_E$ and $V_E$, and matrices $\Delta_U$ and $\Delta_V$, such
that
\begin{eqnarray}
  U_{QE} & = & (U_Q \otimes U_E) (I + \Delta_U); \\
 \| \Delta_U \| & = & \Delta_{Q:E}(U_Q,U_{QE}); \\
  V_{QE} & = & (I + \Delta_V) (V_Q \otimes V_E) ; \\
 \| \Delta_V \| & = &\Delta_{Q:E}(V_Q,V_{QE}). 
\end{eqnarray}
With these choices, $I+\Delta_U$ and $I+\Delta_V$ are easily verified
to be unitary operations on $QE$.  We see that
\begin{eqnarray}
  V_{QE}U_{QE} & = & (I+\Delta_V)(V_Q \otimes V_E) (U_Q \otimes U_E)
  (I+\Delta_U) \nonumber \\
 & & \\
  & = & (I+\Delta_V)(U_Q \otimes V_E) (V_Q \otimes U_E)
  (I+\Delta_U), \nonumber \\
 & &
\end{eqnarray}
where we used the commutativity of $U_Q$ and $V_Q$ in the second line.
The proof is completed by choosing $\tilde U_{QE} \equiv
(I+\Delta_V)(U_Q \otimes V_E)$ and $\tilde V_{QE} \equiv (V_Q \otimes
U_E)(I+\Delta_U)$.

\textbf{QED}

%
% noise on subspaces
%
To prove our threshold theorems for cluster-state computation, we need
two other theorems, which we call the first and second unitary
extension theorems.  These results are not phrased directly in terms
of the error strength $\Delta_{Q:E}$, but we shall see later in the
paper that these theorems have significant implications for the error
strength.

%
% first uet
%
The first unitary extension theorem may be motivated by the following
problem.  Suppose $V$ is a noisy unitary operation approximating a
noiseless unitary operation $U$.  (Note that $U$ and $V$ here act on
the same state space; $U$ might correspond to $U_Q \otimes U_E$ in our
earlier notation, with $V$ corresponding to $V_{QE}$.)  For some
physical reason we are only interested in the action of $U$ and $V$ on
some subspace $S$ of the total Hilbert space.  That is, we know on
physical grounds that all inputs to the operations are constrained to
be in that subspace.  Furthermore, there is another unitary operation
$\tilde U$ which acts \emph{identically} to $U$ on the subspace $S$.
A natural question is whether we can find a unitary operation $\tilde
V$ which acts identically to $V$ on the subspace $S$, and so that
$\tilde V$ approximates $\tilde U$ at least as well as $V$
approximates $U$.

Remarkably, such an extension $\tilde V$ always exists, and the proof
of the first unitary extension theorem shows how it may be
constructed.  We believe this theorem has considerable independent
interest in its own right, quite apart from the applications later in
this paper to fault-tolerant computation with cluster states.

\begin{theorem} {} \label{thm:uet1}
 \textbf{(First unitary extension theorem)}
  Let $U, \tilde U$ and $V$ be unitaries acting on a Hilbert space
  $T$.  Suppose $S$ is a subspace of $T$ such that $U$ and $\tilde U$
  have the same action on $S$, i.e., $U|_S = \tilde U|_S$.  (Note that
  we do not assume that $U$ and $\tilde U$ leave the subspace $S$
  invariant, so $U|_S$ and $\tilde U|_S$ should be considered as maps
  from $S$ into $T$.)  Then there exists a unitary extension $\tilde
  V$ of $V|_S$ to the entire space $T$ such that
  \begin{eqnarray}
    \| \tilde V - \tilde U \| \leq \| V-U \|.
  \end{eqnarray}
\end{theorem}

The proof of this theorem is given in
Appendix~\ref{app:unitary-extension}.  

The second unitary extension theorem answers a question similar in
spirit, but not identical, to the question answered by the first
unitary extension theorem.  Let $U$ and $V$ be unitary operations
acting on a vector space $T$, with a subspace $S$.  Suppose $U|_S$ and
$V|_S$ are close, i.e., $\| U|_S - V|_S\|$ is small.  Can we argue
that there exists a unitary operation $\tilde V$ extending $V|_S$, and
such that $\| U - \tilde V\|$ is also small?  The second unitary
extension theorem shows that this is always true.

\begin{theorem} {} \label{thm:uet2} \textbf{(Second unitary extension theorem)}
  Let $U$ and $V$ be unitary operation acting on a
  (finite-dimensional) inner product space $T$.  Suppose $S$ is a
  subspace of $T$.  Then there exists a unitary operation $\tilde V$
  such that $\tilde V|_S = V|_S$ and
  \begin{eqnarray}
    \| U - \tilde V\| \leq 2 \| U|_S - V|_S\|.
  \end{eqnarray}
\end{theorem}

%
% general comments
%
The proof of the second unitary extension theorem is given in
Appendix~\ref{app:unitary-extension}.  Note that this theorem may
easily be restated in the language of isometries, if that is more to
one's taste.  It is also worth noting that the second unitary
extension theorem implies a weaker version of the first unitary
extension theorem.  In the notation of the first theorem, the second
theorem implies that there exists a unitary extension $\tilde V$ of
$V|_S$ such that $\|\tilde V - \tilde U\| \leq 2 \| \tilde U|_S - V|_S
\| \leq 2 \| U-V\|$.

\subsection{Fault-tolerance in the quantum circuit model}
\label{subsec:circuit-threshold}

The threshold for cluster-state computation proved in this paper is
based on the threshold for quantum circuits proved by Terhal and
Burkard~\cite{Terhal04a}.  In this subsection we review Terhal and
Burkard's result.  We begin with a description of the assumptions they
make about quantum circuits, including the noise model, before stating
their main theorem.  Note that the noise model used by Terhal and
Burkard is the basis for our noise model for cluster-state
computation, described in Subsection~\ref{subsec:noise-model}.

Terhal and Burkard split the total system up into three types of
subsystem.  First, there are \emph{register qubits}, which can be
controlled and used for computation.  These qubits are present through
the entirety of the computation.  Second, there are \emph{ancilla
  qubits}, which may also be controlled and used for the computation.
The difference between register and ancilla qubits is that the
ancillas may be brought into the computer partway through a
computation, used as part of the subsequent computation, and then
discarded at some later time.  The third type of system is the
\emph{environment}, which is not under control.

The computation is represented by a sequence of unitary operations.
Ideally, these operations would be applied just to the register and
ancilla qubits, but inevitably they involve some interaction with the
uncontrolled environment.  It is this interaction which causes noise
in the computer.  We will find it convenient to assume the interaction
with the environment is unitary; by making the environment
sufficiently inclusive the laws of quantum mechanics ensure we may
always make such an assumption.

%
% most important assumption
%
Within this framework, our noise model may be described as follows.
Each qubit in the computer, whether a register qubit or an ancilla
qubit, has associated with it its own environment.  So, for example,
if we label the qubits $Q_1, Q_2,\ldots$, then the corresponding
environments would be labeled $E_1,E_2,\ldots$.  The key assumption
we make about noise is that \emph{non-interacting qubits have
  non-interacting environments}.  More precisely, suppose as part of
the computation we want to attempt some unitary gate $U$ on qubit
$Q_j$.  This might be the identity gate, representing quantum memory,
or it might be a more complex gate, like a Hadamard or Pauli gate.  In
reality, this gate will be noisy, due to interactions with the
environment.  Our assumption is that the real noisy operation is a
unitary evolution $V$ acting on $Q_j$ and its environment $E_j$, with
the other qubits and their environments not affected.  In a similar
way, if we attempt a two-qubit operation $U$ between qubits $Q_j$ and
$Q_k$, we assume that the real noisy evolution $V$ may involve the
qubits $Q_j, Q_k$, and the corresponding environments $E_j, E_k$, but
not the other qubits or their environments.  With these assumptions,
we say the noise in a noisy circuit is of strength at most $\eta$ if
each ideal gate $U_j$ in the circuit is approximated by a noisy gate
$V_j$ such that $\Delta_{Q:E_Q}(U_j,V_j) \leq \eta$, where $Q$ is the
qubit or qubits involved in the gate, and $E_Q$ is the corresponding
environment or environments.

%
% give it a name
%
We refer to the assumption that non-interacting qubits have
non-interacting environments as the \emph{locality assumption} for
noise\footnote{Terhal and Burkard consider even more general noise
  models, which may be of interest in certain circumstances.  However,
  the locality assumption is sufficiently strong to cover a very wide
  class of physically interesting noise models, and so we restrict
  attention to noise models satisfying this assumption.}.  Physically,
the motivation for the locality assumption is that each environment is
well-localized in space, and that environments can only interact with
one another when two qubits are brought together to interact in a
quantum gate.

%
% lack of a Markov approximation
%
Importantly, Terhal and Burkard do not make any Markovian assumption.
That is, each environment can have an arbitrarily long memory. So, for
example, we may perform a sequence of gates in which $Q_1$ first
interacts with $E_1$, which then passes information onto $E_2$ through
a subsequent gate, then onto $E_3$ through another gate, and finally
corrupts qubit $Q_4$, say.  This is in contrast with many other
variants of the threshold theorem, where Markovian noise is assumed,
i.e., qubits are assumed to have independent and memoryless
environments.

%
% additional assumptions
%
In addition to the locality assumption for noise, Terhal and Burkard
make three important additional assumptions about how quantum circuits
are performed:
\begin{enumerate}

\item It is possible to perform quantum gates on different qubits in
  parallel.  Physically, this requirement is due to the fact that
  error-correction must constantly be performed on all the qubits,
  even if one is merely attempting to maintain them in memory.  It is
  possible to prove that parallelizability is a necessary condition
  for a threshold theorem to apply.

\item It is possible to initialize fresh ancilla qubits in the state
  $|0\rangle$ just prior to their being brought into the computation.
  Physically, this requirement is due to the fact that the ancillas
  are used as an entropy sink to remove noise from the computation.
  To be effective in this capacity they must start in a low-entropy
  state.  It is possible to prove that the requirement for fresh
  ancillas is a necessary condition for a threshold theorem to
  apply~\cite{Aharonov96a}.

\item Excepting ancilla preparation, all dynamical operations applied
  during the computation are unitary, up until the final measurement
  at the end of the computation.  This is not a necessary feature of a
  threshold theorem, but is a feature of the threshold of Terhal and
  Burkard.
\end{enumerate}

%
% comment on unitarity
%
The third assumption, that the computation is performed using only
unitary operations, is rather inconvenient from our point of view,
since the cluster-state model of quantum computation inherently
involves many measurements performed during the computation.  One
feature of our proof is that it involves the replacement of
measurements and classical feedforward by equivalent unitary
operations.  The reason we use the all-unitary model is that we need a
threshold theorem which allows non-Markovian noise, and at present
this means using Terhal and Burkard's all-unitary model.  Future
improvements to the threshold theorem for cluster states may come by
developing threshold results for quantum circuits which allow both
non-Markovian noise and measurement during the computation.

%
% locations
%
To conclude preparation for the statement of the threshold theorem we
need a few final items of notation and nomenclature.  It will be
convenient to assume that each unitary operation performed during the
computation takes the same amount of time, $\Delta t$, and so the
circuit may be written as a sequence of unitary operations performed
at times $t = 0, t = \Delta t, t = 2 \Delta t$, and so on.  We define
a \emph{location} in the circuit to be specified by a triple $(k\Delta
t, U, Q)$ consisting of the time $k \Delta t$ at which the gate $U$ is
performed on a qubit or ordered pair of qubits, $Q$.  Our measure of
the total size of the circuit is the total number of locations in that
circuit.  Note that it is important to count locations at which the
identity gate is applied to a qubit.

%
% conclusion of the computation
%
Computation is concluded by measuring the computer in the
computational basis to produce a probability distribution $p$.  The
goal of fault-tolerance is to take a perfect quantum circuit which
outputs a probability distribution $p$ and to construct a
fault-tolerant quantum circuit that may be subject to noise, but
nonetheless outputs a probability distribution $p'$ which is
\emph{close} to $p$ in some suitably defined sense.  As the measure of
closeness we use the \emph{Kolmogorov distance}, $\| p - p' \|_1
\equiv \frac 12 \sum_ x |p(x) - p'(x)|$.

%
% threshold theorem
%
\begin{theorem} {} \textbf{(Threshold theorem for quantum circuits~\cite{Terhal04a})}
  There exists a constant threshold $\eta_{\rm th} > 0$ for quantum
  circuit computation with the following property.  Let $n$ be the
  number of locations in a perfect quantum circuit, $C$, which outputs
  the probability distribution $p$.  Let $\epsilon > 0$.  We can
  efficiently construct a noisy quantum circuit, $C'$, with a total
  number of locations $n \mbox{polylog}(n^2/\epsilon)$, and such that
  if the noise in $C'$ satisfies the locality assumption and is of
  strength at most $\eta_{\rm th}$ then the output distribution $p'$
  from $C'$ satisfies $\| p-p'\|_1 \leq \epsilon$.
\end{theorem}

%
% comment on the set of operations
%
Terhal and Burkard's construction of $C'$ is based on a particular
type of quantum error-correcting code dubbed a \emph{computation code}
by Aharonov and Ben-Or (definition number 15 in~\cite{Aharonov99a}).
As a consequence of this construction, the circuit $C'$ is built up
out of a special restricted class of quantum gates, gates that can be
implemented in a fault-tolerant manner.  For example, it is possible
to construct $C'$ using just operations from the following restricted
set: preparation of qubits in the state $|0\rangle$; the identity
gate, i.e., quantum memory; $H$ (Hadamard) gates, $Z_{\pi/4}$ and
$Z_{\pi/8}$ gates, where $Z_\theta$ is the rotation of a single qubit
by $\theta$ about the $z$ axis of the Bloch sphere; and
controlled-{\sc not} gates.  As noted earlier, $C'$ does not include
any measurement or classical processing of data, except at the output;
all dynamical operations are fully unitary.

%
% replacement circuit
%
For our purposes in this paper it is convenient to replace $C'$ with
an equivalent circuit, $C'''$, built up from a different set of basic
operations.  We make this replacement in two stages.  The first stage
is to replace the operations in the circuit $C'$ by operations from
the following set: preparations of a qubit in the state $|+\rangle$;
the identity gate; gates of the form $X_\alpha Z_\beta$; and the
controlled-$Z$ gate, which we shall call {\sc cphase}.  That this can
always be done follows from well-known quantum circuit identities.  We
call the resulting circuit $C''$.

%
% second stage
%
The second stage is to replace the operations in $C''$ by operations
from the following set: preparations of a qubit in the state
$|+\rangle$; gates of the form $HZ_{\alpha}$; and the gate $(H\otimes
H)${\sc cphase}.  We refer to this as the \emph{canonical set} of
allowed operations.  To see that this can be done requires a little
care, due to the absence of the identity gate from the canonical set.
The trick is to simulate each gate in $C''$ by two gates from the
canonical set, as follows: $I \rightarrow H H$; $X_{\alpha} Z_{\beta}
\rightarrow H Z_{\alpha} H Z_{\beta}$; {\sc cphase}$\rightarrow (H
\otimes H) (H \otimes H)${\sc cphase}.

%
% canonical form
%
We call the circuit that results when these substitutions are made
$C'''$, and refer to it as the \emph{canonical form} of the
fault-tolerant circuit $C'$.  It is clear on physical grounds that the
canonical form also satisfies the threshold theorem.  Alternately, a
rigorous proof of this fact follows from the chaining property for
error strength, Proposition~\ref{prop:chaining}.  The essential idea
of the proof may be illustrated by example: suppose an identity gate
in the original circuit $C'$ has been replaced by two consecutive $H$
gates in the canonical circuit $C'''$.  Provided the $H$ gates both
suffer from noise of strength less than $\eta_{\rm th}/2$,
Proposition~\ref{prop:chaining} ensures that their product is
equivalent to doing the identity gate with error strength at most
$\eta_{\rm th}$.  Thus, while the threshold $\eta_{\rm th}'''$ for
$C'''$ may be somewhat reduced from the threshold for $C'$, $\eta_{\rm
  th}$, it is reduced at most by some constant factor.

%
% summing up
%
Summing up, we have the following restatement of the threshold theorem
in the form that will be used for our analysis of fault-tolerant
cluster-state quantum computation.

\begin{theorem} {} \textbf{(Threshold theorem for quantum computation,
    with circuits in canonical form)} \label{thm:threshold-canonical}
  There exists a constant threshold $\eta_{\rm th} > 0$ for quantum
  circuit computation with the following property.  Let $n$ be the
  number of locations in a perfect quantum circuit, $C$, which outputs
  the probability distribution $p$.  Let $\epsilon > 0$.  We can
  efficiently construct a noisy quantum circuit, $C'$, using only
  operations from the canonical set of operations (preparation of a
  qubit in the state $|+\rangle$; gates of the form $HZ_{\alpha}$; and
  the gate $(H\otimes H)$ {\sc cphase}), with a total number of
  locations $n \mbox{polylog}(n^2/\epsilon)$, and such that if the
  noise in $C'$ satisfies the locality assumption and is of strength
  at most $\eta_{\rm th}$ then the output distribution $p'$ from $C'$
  satisfies $\| p-p'\|_1 \leq \epsilon$.
\end{theorem}

%% file: ftqc-cs-csc.tex
\section{Cluster-state quantum computation}
\label{sec:cluster-state-qc}

%
% what we do
%
In this section we describe how cluster-state quantum computation
works.  Subsec.~\ref{subsec:basic-cluster-state} gives a basic
description of the model, and introduces language useful in the later
analysis of fault-tolerance.
Subsec.~\ref{subsec:optical-cluster-state} describes how cluster-state
computation can be realized in optics.  All proofs are omitted, and
the reader is referred instead to~\cite{Raussendorf01a}, or to the
leisurely pedagogical account in~\cite{Nielsen03c}.  Note that our
account barely scratches the surface of the work that has been done on
cluster-state computation: the interested reader should also
consult~\cite{Raussendorf02a,Raussendorf03a,Aliferis04a,Jorrand04a,Childs04a}
for further work on the cluster-state model of quantum computation;
further work on other measurement-based models of quantum computation
may be found
in~\cite{Fenner01a,Leung01c,Leung03a,Jorrand03a,Perdrix04b,Perdrix04a,Perdrix04c}.

\subsection{Introduction to cluster-state quantum computation}
\label{subsec:basic-cluster-state}

\begin{figure}[ht]
\scalebox{0.95}{\epsfig{file=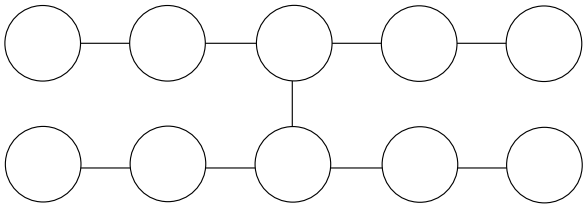}}

\caption{A simple cluster state.  Note that each circle represents a
  single qubit in the cluster.
\label{fig:basic-cluster-state}}
\end{figure}

%
% define cluster states
%
The basic element of the cluster-state model is the cluster state, an
entangled network of qubits\footnote{The states we call cluster states
  are in fact a generalization of the cluster state used
  in~\cite{Raussendorf01a}. These generalized states have been called
  \emph{graph states} elsewhere; we prefer to use the more elegant
  term cluster state to refer to all the states in this class.}.  An
example of a cluster state is represented in
Figure~\ref{fig:basic-cluster-state}.  Each circle represents a single
qubit in the cluster.  We may define the cluster state as being the
result of the following two-part preparation procedure: first, prepare
each qubit in the state $|+\rangle \equiv (|0\rangle+|1\rangle)/\sqrt
2$, and then apply {\sc cphase} gates between any two qubits joined by
a line.  Since the {\sc cphase} gates commute with one another, it
does not matter in what order they are applied.  Note that this is
merely a convenient way of defining the cluster state, and there is no
requirement that it be prepared in this way.

%
% how a cluster-state computation proceeds
%
Given the cluster state, a cluster-state computation is simply a
procedure for measuring some subset of qubits in the cluster, using
single-qubit measurements and feedforward of the measurement results
to control the bases in which later qubits are measured.  The output
of the computation is the joint state of whatever qubits remain
unmeasured at the end of the computation.

\begin{figure}[ht]
\epsfig{file=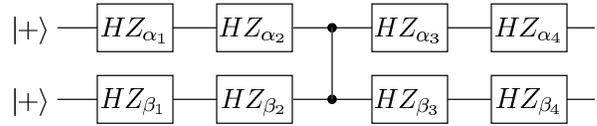}

\caption{A two-qubit quantum circuit. Without loss of generality we
  may assume the computation starts with each qubit in the $|+\rangle
  \equiv (|0\rangle+|1\rangle)/\sqrt{2}$ state, since single-qubit
  gates can be prepended to the circuit if we wish to start in some
  other state.  The two-qubit gate is a {\sc cphase} gate.  The boxes
  are single-qubit gates of the form $HZ_\alpha$, where $H$ and
  $Z_\alpha$ are as defined in
  Subsection~\ref{subsec:circuit-threshold}.  Note that by composing
  three of these gates we can obtain an arbitrary single-qubit gate.
  Thus, gates of the form $HZ_\alpha$, together with {\sc cphase}
  gates, are universal for quantum computation.
 \label{fig:basic-circuit}}
\end{figure}

\begin{figure}[ht]
\scalebox{0.95}{\epsfig{file=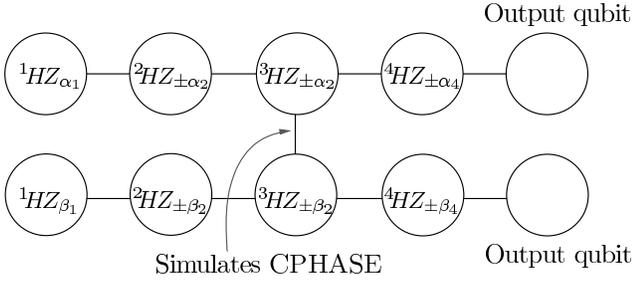}}
\caption{The cluster state of Figure~\ref{fig:basic-cluster-state},
  marked to indicate how it is used to simulate the different elements
  in the circuit of Figure~\ref{fig:basic-circuit}.  Note that the
  labeled qubits all have labels of the form ``$^nU$'', where $n$ is
  a positive integer, and $U$ is a unitary operation.  The label $n$
  indicates the time order; qubits with the same label can be measured
  in either order, or simultaneously.  $U$ indicates that the qubit is
  measured by performing the unitary $U$ and then measuring in the
  computational basis.  Equivalently, a single-qubit measurement in
  the basis $\{ U^\dagger |0\rangle, U^\dagger |1\rangle\}$ is
  performed.  Note that except for the first measurements, all
  measurements have $U$ of the form $HZ_{\pm \alpha}$; the $\pm$
  indicates that the value of the sign depends upon the outcomes of
  earlier measurements.  The output from the computation is at the
  unlabeled qubits, which are not measured.
\label{fig:correspondence-cluster-state}}
\end{figure}

%
% explain how this can be used to simulate a quantum circuit
%
Remarkably, this procedure can be used to simulate an arbitrary
quantum circuit.  Indeed, the cluster state of
Figure~\ref{fig:basic-cluster-state} was specifically chosen in order
to simulate the circuit in Figure~\ref{fig:basic-circuit}. In
Figure~\ref{fig:correspondence-cluster-state} we illustrate visually
how the cluster state of Figure~\ref{fig:basic-cluster-state} can be
used to simulate the circuit in Figure~\ref{fig:basic-circuit}.  Each
qubit in the quantum circuit is replaced by a horizontal line of
qubits in the cluster state.  Different horizontal qubits in the
cluster represent the original qubit at different times, with the
progress of time being from left to right.  Each single-qubit gate $H
Z_{\alpha}$ in the quantum circuit is replaced by a single qubit in
the cluster state.  {\sc cphase} gates in the original circuit are
simulated using a vertical ``bridge'' connecting the appropriate
qubits.  The cluster-state computation itself is carried out by
performing a series of measurements in the time order and measurement
bases indicated in the caption to
Figure~\ref{fig:correspondence-cluster-state}.  The final output of
the cluster-state computation $|\psi\rangle$ is related to the output
of the quantum circuit $|\phi\rangle$ by $|\psi\rangle = \sigma
|\phi\rangle$, where $\sigma$ is a product of Pauli matrices that is
an easy-to-compute function of the measurement outcomes obtained
during the cluster-state computation.

Although the example we have described involves a specific quantum
circuit, general quantum circuits can be given a cluster-state
simulation along similar lines.  Note that we have not explicitly
explained how the measurement feedforward procedure works, nor the
precise function of measurement outcomes that determines the Pauli
correction $\sigma$ at the end of the computation.  These are
explained in detail in~\cite{Raussendorf01a,Nielsen03c}; we also give
an explicit description of the procedures used in
Section~\ref{sec:deterministic}.

\begin{figure}[ht]
\epsfig{file=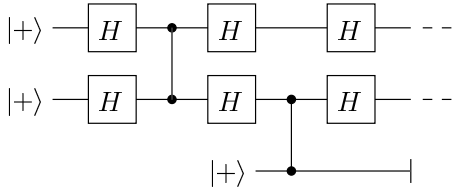} \caption{A simple quantum
circuit in which an ancilla is prepared and later discarded.
 \label{fig:circuit-with-ancilla}}
\end{figure}

\begin{figure}[hb]
\epsfig{file=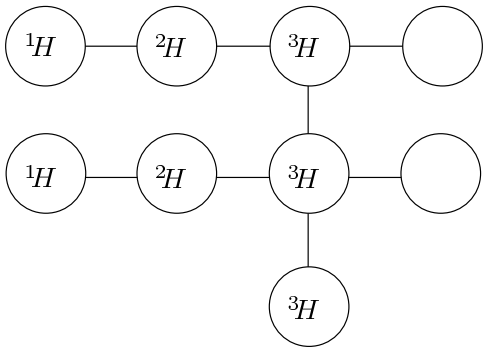} \caption{The
cluster-state computation used to simulate the circuit in
  Figure~\ref{fig:circuit-with-ancilla}.
 \label{fig:cluster-ancilla-example}}
\end{figure}

One feature of our example quantum circuit,
Figure~\ref{fig:basic-circuit}, that deserves attention is the fact that
it doesn't involve any ancilla qubits.  An important feature of many
quantum circuits is that they involve the preparation and discarding
of ancillas during the computation.  This is especially true of
circuits for quantum error-correction and fault-tolerant quantum
computation, where the ancillas are used as a heat sink to remove
excess entropy from the computer.  Such ancilla preparations and
removal are easily simulated in the cluster-state model.
Figure~\ref{fig:circuit-with-ancilla} illustrates a simple quantum
circuit involving an ancilla that is prepared and later discarded.
Figure~\ref{fig:cluster-ancilla-example} illustrates how this
preparation and discarding can be simulated within the cluster-state
model.

The cluster states we have described so far have all been embedded in
two dimensions.  This is for convenience only.  In practice, a more
complicated topology for the cluster may be useful in some
circumstances.  This may be achieved either by embedding the cluster
in a higher number of dimensions, or by nonlocal connections between
different parts of the cluster.  Fault-tolerant quantum circuits often
involve two (or more) spatial dimensions, corresponding to a
three-dimensional cluster-state computation.  Note, however, that it
does \emph{not} follow that we require the use of all three spatial
dimensions to do a cluster-state simulation of a two-dimensional
quantum circuit.  We will see below that it is only necessary to
prepare a small part of the cluster at any given time, and this means
that the cluster-state computation may be performed without requiring
additional spatial dimensions beyond those used in the circuit being
simulated.

%
% simplify the discussion to two dimensions
%
Despite the different possible topologies of the cluster state, when
convenient we shall continue to discuss cluster states as though they
are embedded in two dimensions.  This considerably simplifies
discussion (and the figures), and the extensions to more complicated
cluster topologies are in all cases obvious.

%
% nomenclature
%
To facilitate later discussion of fault-tolerance we now introduce
some additional nomenclature to describe cluster-state computations.
We call a single vertical slice of qubits through the cluster a
\emph{layer} of the cluster.  For example, in
Figure~\ref{fig:correspondence-cluster-state} the two qubits with the
label $4$ form a layer.  A cluster-state simulation of a circuit thus
consists of performing single-qubit measurements, layer by layer.  We
can think of a single layer in a cluster-state computation as similar
to the instantaneous quantum state at some fixed time during a quantum
circuit computation.

%
% levels
%
We call a single horizontal row of qubits through the cluster a
\emph{level}.  So, for example, all the qubits on the top line of
Figure~\ref{fig:correspondence-cluster-state} represent a level.  We
can think of a level as representing the evolution of a single qubit
in a quantum circuit computation.  We call a cluster-state computation
involving only a single level a \emph{single-qubit} cluster-state
computation.  The motivation for this nomenclature is that, as we have
seen, such cluster-state computations are used to simulate
single-qubit quantum circuits.  Note that with this definition a
single-qubit cluster-state computation may involve more than one
physical qubit.  A multi-qubit cluster-state computation is one that
involves more than a single level of qubits.

%
% implementation
%
Up until now, we have described a cluster-state computation as being
composed of two steps: preparation of the cluster state, followed by
an adaptive sequence of single-qubit measurements.  However, it is
also possible to implement cluster-state computations in alternative
ways.  The key observation is that we can delay preparation of some
parts of the cluster until later, doing some of the measurements
first.  So, for example we could do a cluster-state computation via
the following sequence of steps:
\begin{itemize}
\item Prepare layers one and two of the cluster, using $|+\rangle$
  preparations and {\sc cphase} gates.
\item Perform the first layer of measurements.
\item Prepare layer three, using $|+\rangle$ preparations and {\sc
    cphase} gates to adjoin the third layer of qubits to the second
  layer.
\item Perform the second layer of measurements.
\item Keep alternating the steps of preparing an extra layer then
  measuring an extra layer, until the end of the computation.
\end{itemize}
We call this a \emph{one-buffered implementation} of cluster state
computation, since there is always a buffer of one layer between the
layer of qubits being measured, and the most recently prepared layer
of qubits.  We call the set of qubits about to be measured the
\emph{current layer}, and the layer after that the \emph{next layer}.

%
% connection to fault-tolerance
%
For fault-tolerance the one-buffered implementation of cluster-state
computation has a great advantage over our original prescription in
which the entire cluster is prepared first.  The reason, as indicated
in the introduction, is that if the entire cluster is prepared first,
then qubits which are to be measured later in the computation will
have undergone substantial degradation by the time they are measured,
and this will unacceptably corrupt the output of the
computation\footnote{We thank Andrew Childs and Debbie Leung for
  pointing this fact out.}.

More generally, the one-buffered implementation illustrates the
important point that a given cluster-state computation may have many
different \emph{implementations}, i.e., different methods for creating
the cluster and performing the required single-qubit measurements.
When proving fault-tolerant threshold theorems for cluster-state
computation we will need to carefully specify the details of the
implementation used.  All the implementations used in this paper are
variants on the one-buffered implementation.

\subsection{Optical cluster-state quantum computing}
\label{subsec:optical-cluster-state}

%
% segue into optical cluster states
%
Our description of cluster-state computation has been as an abstract
model of quantum computation.  As described in the introduction the
cluster-state model also shows great promise as the basis for
experimental implementations of quantum computation in
optics~\cite{Nielsen04b}.  We now briefly describe the optical
implementation of cluster-state computation,
following~\cite{Nielsen04b}, and some of the special challenges it
poses for a proof of fault-tolerance.

%
% relate to KLM
%
The proposal of~\cite{Nielsen04b} is a modified version of the
proposal of Knill, Laflamme and Milburn (KLM)~\cite{Knill01a}, and we
now briefly review some of the basic elements of KLM, following the
review in~\cite{Nielsen04b}.  KLM encodes a single qubit in two
optical modes, $A$ and $B$, with logical qubit states $|0\rangle_L
\equiv |01\rangle_{AB}$, and $|1\rangle_L \equiv |10\rangle_{AB}$.
(Note that we are using the standard Bosonic occupation number
representation on the right-hand side of these definitions, not the
qubit notation, so that, for example, $|01\rangle_{AB}$ indicates zero
photons in mode $A$, and one photon in mode $B$.)  State preparation
is done using single-photon sources, while measurements in the
computational basis may be achieved using high-efficiency
photodetectors.  Such sources and detectors make heavy demands not
entirely met by existing optical technology, although encouraging
progress on both fronts has been reported recently.  Arbitrary
single-qubit operations are achieved using phase shifters and
beamsplitters.

%
% the non-det dest gate
%
The main difficulty in KLM is achieving near-deterministic entangling
interactions between qubits.  KLM use the idea of gate
teleportation~\cite{Gottesman99a,Nielsen97c} to produce a gate
$CZ_{n^2/(n+1)^2}$ which with probability $n^2/(n+1)^2$ applies a {\sc
  cphase} to two input qubits, where $n$ is any fixed positive
integer.  When the gate fails, the effect is to perform a measurement
of those qubits in the computational basis.  Increasing values of $n$
correspond to increasingly complicated teleportation circuits.  For
this reason, KLM combine these ideas with ideas from quantum
error-correction in order to achieve a near-deterministic {\sc cphase}
gate, and thus complete the set required for universal quantum
computation.

%
% introduce the term ``postselected''
%
An important property of the gate $CZ_{n^2/(n+1)^2}$ is that in the
ideal case of perfect implementation, \emph{we know when the gate
  succeeds}.  In particular, in KLM's implementation procedure,
success of the gate is indicated by certain photodetectors going
``click'', while failure is indicated by different photodetection
outcomes.  We call such gates \emph{postselected} gates to indicate
that whether the gate has succeeded is known, and can be fed forward
to later parts of the computation.

%
% idea of optical cluster states
%
The advantage of the optical cluster-state proposal
of~\cite{Nielsen04b} is that it only makes use of the $CZ_{1/4}$ and
$CZ_{4/9}$ gates, both of which use relatively simple configurations
of optical elements, and avoids the use of error-correction in
achieving a near-deterministic {\sc cphase} gate.  This results in a
greatly simplified proposal for quantum computation.

%
% idea of OCSC
%
A key observation used in the optical cluster-state proposal is an
interesting general property of cluster states. Suppose we measure one
of the cluster qubits in the computational basis, with outcome $m$.
Then it can be shown that the posterior state is just a cluster state
with that node deleted, up to a local $Z^m$ operation applied to each
qubit neighbouring the deleted qubit.  These are known local
unitaries, whose effect may be compensated in subsequent operations,
so we may effectively regard such a computational basis measurement as
simply removing the qubit from the cluster.

%
% how it works
%
This is a useful observation because when the $CZ_{n^2/(n+1)^2}$ gate
fails, it effects a measurement in the computational basis.  Thus, if
one attempts to add qubits to a cluster using a $CZ_{n^2/(n+1)^2}$
gate, failure of the gate merely results in a single qubit being
removed from the cluster, rather than the entire cluster being
destroyed.  \cite{Nielsen04b} shows that by combining this observation
with a random walk technique, it is possible to efficiently build up
an arbitrary cluster state using either $CZ_{4/9}$ or $CZ_{1/4}$
gates.  Once this is done, all the other operations in the
cluster-state model can be done following KLM's prescription.

%% file: ftqc-cs-deterministic.tex
\section{Fault-tolerance with deterministic $CPHASE$ gates}
\label{sec:deterministic}

In this section we prove a threshold theorem for noisy cluster-state
quantum computation.  This theorem is applicable to situations in which
the cluster can be extended during the computation using {\sc
  cphase} gates that are noisy, but operate deterministically.  In the
next section we extend the theorem to some situations where the {\sc
  cphase} gates operate non-deterministically, as is the case for
optical cluster-state computation.

Rather than attempt to invent fault-tolerant methods for cluster-state
computation from scratch, it is natural to build off the existing and
rather extensive body of literature on fault-tolerant quantum
circuits.  As described in the introduction, our strategy is to
consider a cluster-state computation that simulates a fault-tolerant
quantum circuit, and then ask if the simulated fault-tolerant
capabilities are able to correct noise in the cluster-state
implementation.

We therefore begin with a quantum circuit $\mathcal{Q}$ and, instead
of directly translating it into the cluster-state model, first encode
$\mathcal{Q}$ as a fault-tolerant circuit $\mathcal{F_Q}$ in the
canonical form of Theorem~\ref{thm:threshold-canonical}. Recall that a
canonical fault-tolerant circuit uses only preparations of qubits in
the state $|+\rangle$, single-qubit gates of the form $H Z_\alpha$,
and the two-qubit gate $\left(H \otimes H \right)${\sc cphase}.  Using
the prescription described in Section~\ref{sec:cluster-state-qc} it is
a simple matter to translate $\mathcal{F_Q}$ into a cluster-state
computation, which we denote $\mathcal{C}$.

Suppose now that $\mathcal{C}^\prime$ is a noisy one-buffered
implementation of $\mathcal{C}$. Is $\mathcal{C}^\prime$ equivalent to
some noisy implementation $\mathcal{F_Q}^\prime$ of $\mathcal{F_Q}$?
We will show in this section that this is indeed the case, and
moreover that the noise is of a type and strength that is correctable
by the fault-tolerance built into $\mathcal{F_Q}$.  The noisy
cluster-state computation $\mathcal{C}^\prime$ is therefore a
fault-tolerant simulation of the original quantum circuit
$\mathcal{Q}$.  It is worth noting that this noise correspondence
holds for \emph{any} quantum circuit and its corresponding
one-buffered implementation; we don't use any special properties of
${\cal F_Q}$ in proving the noise correspondence.

% --------------------------------------------------------------------- %
% FIGURE: Single-Qubit Circuit                                          %
\begin{figure} %[ht]
\epsfig{file=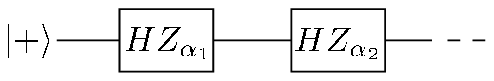}
\caption{A single-qubit circuit containing only operations from
  the canonical set.
\label{fig:single-qubit-circuit}}
\end{figure}
% --------------------------------------------------------------------- %

% --------------------------------------------------------------------- %
% FIGURE: Single-Qubit Cluster State Computation                        %
\begin{figure}[hb]
\includegraphics{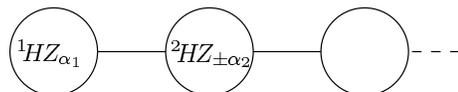}
\caption{A cluster-state computation implementing the
single-qubit circuit shown in Figure~\ref{fig:single-qubit-circuit}. 
We will show that the
noise in this single-qubit cluster-state computation can
 be mapped onto equivalent noise in the corresponding quantum circuit.
\label{fig:single-qubit-csc}}
\end{figure}
% --------------------------------------------------------------------- %

% --------------------------------------------------------------------- %
% FIGURE: Literal circuit of the single-qubit CSC                       %
\begin{figure} %[ht]
\includegraphics{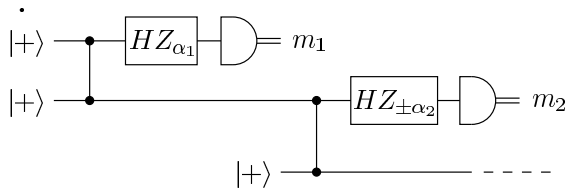}
\caption{The literal circuit for the one-buffered implementation of the
cluster-state computation in Figure~\ref{fig:single-qubit-csc}.
\label{fig:sq-csc-onebuff}}
\end{figure}
% --------------------------------------------------------------------- %

A key construction used in establishing this noise correspondence is
what we call the \emph{literal} quantum circuit, $\mathcal{L}$.  The
literal circuit is a quantum circuit depiction of the operations
performed during a one-buffered implementation of a cluster-state
computation. It is a literal translation of the one-buffered
implementation, and should not be confused with the quantum circuit
$\mathcal{F_Q}$ being simulated.  As an example, consider the
single-qubit quantum circuit depicted in
Figure~\ref{fig:single-qubit-circuit}.  The corresponding
cluster-state computation is depicted in
Figure~\ref{fig:single-qubit-csc}, and the literal circuit showing the
one-buffered implementation is shown in
Figure~\ref{fig:sq-csc-onebuff}. Note that although $3$ qubits appear
in the literal circuit, the quantum circuit being simulated is a
single-qubit computation.

The literal circuit $\mathcal{L}$ offers a convenient means for
describing the effects of noise in $\mathcal{C}^\prime$, and for this
reason we have gone to some trouble in Figure~\ref{fig:sq-csc-onebuff}
to depict the correct time-ordering of events. We have, for example,
offset the preparation of the final $|+\rangle$ state, since it is not
actually prepared until later in the one-buffered implementation, and
preparation at an earlier time would result in considerably more noise
affecting the qubit.

The noise in $\mathcal{C}^\prime$ is quantified by the error strength
$\Delta_{Q:E}$ of the operations appearing in the literal circuit.
The key result of this section is that if the worst-case noise
strength in $\mathcal{C}^\prime$ is $\eta$, then the corresponding
noise in the quantum circuit $\mathcal{F_Q}^\prime$ satisfies the
locality assumption and has strength at most $c\eta$, for some
constant $c$.  Provided $c\eta \leq \eta_\mathrm{th}$ we conclude that
the distribution $p'$ that results when we measure the output of the
cluster-state computation satisfies $\| p-p'\|_1 \leq \epsilon$, where
$p$ is the distribution output from the noise-free computation.

The section contains three parts. Subsection~\ref{subsec:noise-model}
introduces our noise model for cluster-state computation.
Subsection~\ref{subsec:single-qubit-step} proves the noise
correspondence described above for the simplest case when the quantum
circuit $\mathcal{F_Q}$ and the corresponding cluster-state
computation $\mathcal{C}$ are single-qubit computations. All the ideas
introduced in this subsection are then extended to the case of a
multi-qubit $\mathcal{F_Q}$ and $\mathcal{C}$ in
Subsection~\ref{subsec:multi-qubit-step}.

% -------------------------------------------------------------------- %
% SECTION: NOISE MODEL
% -------------------------------------------------------------------- %

\subsection{Noise model for a one-buffered cluster-state computation}
\label{subsec:noise-model}

The noise model appropriate to a cluster-state computation depends
critically upon the implementation procedure used to perform the
computation.  The results in this section are based on the
one-buffered implementation procedure, as described in
Section~\ref{sec:cluster-state-qc}.  Recall that in a one-buffered
implementation the only qubits available in the cluster at any given
time are the qubits in the current layer, and the next layer.  The
computation is performed by repeatedly performing the following two
steps: (1) making all the necessary measurements on the current layer;
and (2) using {\sc cphase} gates to add an extra layer of qubits into
the cluster.  The only variation in this procedure comes at the very
beginning of the computation, where we need to create two whole layers
of cluster-state qubits, and at the end, where we don't need to add an
extra layer into the cluster.

As in the fault-tolerance results for quantum circuits, our results do
not allow for completely arbitrary types of noise.  Instead, we make
some physically plausible assumptions about the nature of noise in the
one-buffered implementation. The noise model we adopt allows for the
following types of noise:

\begin{enumerate}

\item \emph{Noise in unitary dynamics:} We model this in a manner
  similar to the noise model for quantum circuits described in
  Subsection~\ref{subsec:circuit-threshold}.  Each qubit has its own
  environment, and we assume that non-interacting qubits have
  non-interacting environments, but make no other assumptions about
  the noise.  Indeed, we can use an even more general noise model, in
  which qubits at the same level in the cluster state are assumed to
  share a common environment, and all we assume is that
  non-interacting levels have non-interacting environments.
  
\item \emph{Noise in quantum memory:} Quantum memory is simply the
  (unitary) identity operation, and we model a noisy quantum memory
  step as we would any other noisy unitary operation.  Note that this
  type of noise affects all qubits other than the current layer during
  a round of measurements.

\item \emph{Noise in preparation of the $|+\rangle$ state:} We model
  this is as perfect preparation of $|+\rangle$, followed by a noisy
  quantum memory step.

\item \emph{Noise in measurements in the computational basis:} We model
  this as a noisy quantum memory step, followed by a perfect
  measurement in the computational basis.
\end{enumerate}
We quantify the overall strength of noise in a one-buffered
cluster-state computation by the worst-case error strength in any of
the unitary operations, including the noisy quantum memory steps in
preparation and measurement.

% -------------------------------------------------------------------- %
% SINGLE-QUBIT NOISE CORRESPONDENCE
% -------------------------------------------------------------------- %

\subsection{Noise correspondence for single-qubit computations}
\label{subsec:single-qubit-step}

In this section we consider the simplest case of a single-qubit
circuit $\mathcal{F_Q}$ made up of gates $HZ_{\alpha_j}$, as shown in
Figure~\ref{fig:single-qubit-circuit}. A cluster-state computation
$\mathcal{C}$ simulating $\mathcal{F_Q}$ is shown in
Figure~\ref{fig:single-qubit-csc}. The establishment of the noise
correspondence is a five-step process.  To help orient the reader, we
now outline these steps.  Note that the meaning of these steps may not
be completely clear upon a first read, but hopefully will ease
comprehension of later parts of the paper.

\begin{figure} %[ht]
\scalebox{0.9}{ \epsfig{file=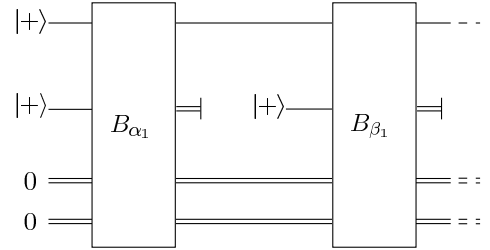}} \caption{The output
of this imperfect circuit is the same as the output of the noisy
cluster-state computation in Figure~\ref{fig:sq-csc-onebuff}.
\label{fig:sq-csc-b}}
\end{figure}

\begin{figure} %[hb]
\scalebox{0.8}{ \epsfig{file=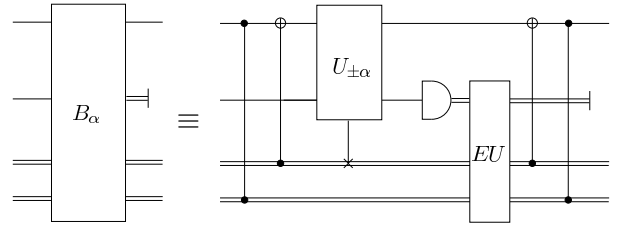}} \caption{The definition
of the operation $B_\alpha$ used as the basis for the repeating blocks in
Figure~\ref{fig:sq-csc-b}. \label{fig:def-b}}
\end{figure}

\begin{figure} %[ht]
\scalebox{0.85}{\epsfig{file=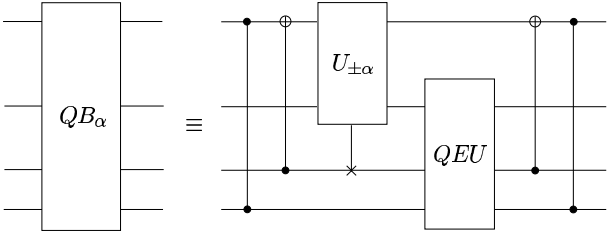}} \caption{The
definition of the operation $QB_\alpha$. \label{fig:def-qb}}
\end{figure}

\begin{figure} %[hb]
  \epsfig{file=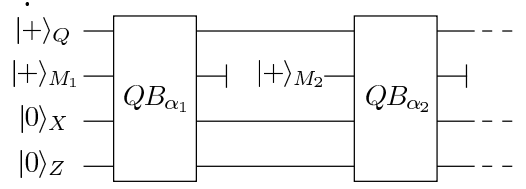} \caption{The output at $Q$ in
    this imperfect circuit is identical to the output of the imperfect
    circuit of Figure~\ref{fig:sq-csc-b}, and thus is equivalent to
    the output of the noisy cluster-state computation of
    Figure~\ref{fig:sq-csc-onebuff}. \label{fig:sq-csc-qb}}
\end{figure}

%
% List the steps
%
\begin{enumerate}
\item We begin with the literal circuit $\mathcal{L}$ depicting a
  one-buffered implementation of $\mathcal{C}$. Such a circuit is
  shown in Figure~\ref{fig:sq-csc-onebuff}.
  
\item The literal circuit of Figure~\ref{fig:sq-csc-onebuff} does not
  explicitly contain the classical feedforward and control that is
  performed during the cluster-state computation.  Without taking
  these into account, it is not possible to understand the effects of
  noise on the computation.  Thus, we expand $\mathcal{L}$ to
  explicitly include the classical feedforward and control.
  
\item We use a series of circuit identities to transform the literal
  circuit $\mathcal{L}$ into an equivalent circuit that contains
  ``block'' operations $B_{\alpha_j}$, each of which correspond
  directly to the action of some gate $H Z_{\alpha_j}$ in
  $\mathcal{F_Q}$.  Looking ahead, the block form equivalent to a
  perfectly implemented $\mathcal{L}$ is depicted in
  Figure~\ref{fig:sq-csc-b}, with the $B_{\alpha_j}$ shown in
  Figure~\ref{fig:def-b}. 
 
\item The threshold theorem of Terhal and Burkard
  involves only unitary operations.  Thus,
  the next step is to replace the classical elements of the blocks
  $B_{\alpha_j}$ with unitary quantum equivalents to obtain a unitary
  operation $Q B_{\alpha_j}$.  Looking ahead, the correspondence with
  the $H Z_{\alpha_j}$ of $\mathcal{F_Q}$ is explicitly shown in
  Proposition~\ref{prop:quantum-circuit-identity}.  The circuit
  equivalent to the literal circuit but containing unitary blocks is
  shown in Figure~\ref{fig:sq-csc-qb}, with the $Q B_{\alpha_j}$
  defined in Figure~\ref{fig:def-qb}.
  
\item It can now be shown that noise in $\mathcal{C}$ is equivalent to
  noise within the unitary blocks $Q B_{\alpha_j}$. Using
  Proposition~\ref{prop:quantum-circuit-identity} and the first
  unitary extension theorem, Theorem~\ref{thm:uet1}, we show that
  noise of strength $\eta$ in $Q B_{\alpha_j}$ corresponds to noise of
  strength at most $c \eta$ in $H Z_{\alpha_j}$, for some appropriate
  constant $c$.

\end{enumerate}

\begin{figure} %[ht]
  \scalebox{0.9}{ \includegraphics{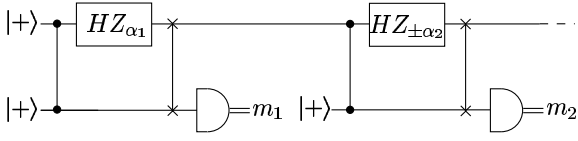} }
  \caption{A circuit with the same output as that of
    Figure~\ref{fig:sq-csc-onebuff}. The new two-qubit gates that have
    been inserted are {\sc swap} gates.
    \label{fig:sq-csc-swap}}
\end{figure}

Our first task in this subsection, then, is to explicitly insert the
classical control and feedforward operations into the literal circuit
$\mathcal{L}$ of Figure~\ref{fig:sq-csc-onebuff}, and to arrange that
circuit into an appropriate block form.  We begin by inserting perfect
{\sc swap} gates into $\mathcal{L}$ to obtain a more compact (but
equivalent) circuit which is shown in Figure~\ref{fig:sq-csc-swap}.
We may assume that these {\sc swap} gates operate perfectly as they are
merely a mathematical convenience. The remaining operations in
Figure~\ref{fig:sq-csc-swap}, however, are real operations and will be
subject to noise. When we need to emphasize that a circuit contains a
mix of both real and perfect operations we will refer to it as an
\emph{imperfect} circuit, as distinct from a \emph{noisy} circuit
where all operations are subject to noise.

% --------------------------------------------------------------------- %
% FIGURE: Insert SWAP into literal circuit                              %
% --------------------------------------------------------------------- %

There are two classical aspects of the cluster-state computation that
we have not yet explicitly included in the literal circuit. These are
the Pauli corrections introduced by the measurement, and the classical
feedforward of measurement results to account for these corrections.
As an example, consider the first measurement in
Figure~\ref{fig:sq-csc-onebuff}. This introduces a correction
$X^{m_1}$ to the state of the qubit immediately below and consequently
the second measurement is performed in the basis $H Z_{\pm \alpha_2}$
according to whether $m_1$ is $0$ or $1$. In general, the Pauli
correction is given by $X^x Z^z$ where $x$ and $z$ are classical
variables. Initially $x$ and $z$ are both zero, and they are updated
after each measurement by the rule
\begin{eqnarray}
  x' & = & z+m\,\,(\mathrm{mod}\,\, 2) \label{eq:EU-1} \\
  z' & = & x\,\,(\mathrm{mod}\,\, 2), \label{eq:EU-2}
\end{eqnarray}
where $m = 0,1$ is the measurement outcome. Subsequent measurement is
performed in the basis $H Z_{\pm \alpha_j}$, with the choice
determined by whether $x$ is $0$ or $1$.

\begin{figure}%[ht]
\scalebox{0.85} {\epsfig{file=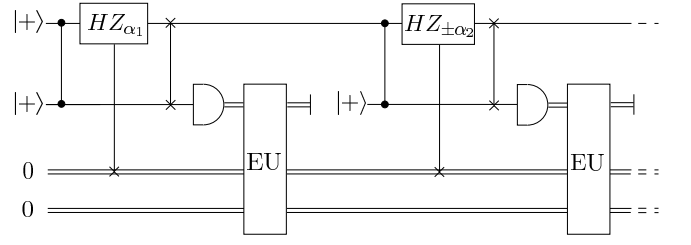}}
\caption{A circuit with the same output as  Figure~\ref{fig:sq-csc-swap},
  but with the classical controls explicitly drawn in.  Note that $x$
  is the top bit, while $z$ is the lower bit.  The new circuit
  notation involving the $HZ_{\alpha_1}$ and $HZ_{\alpha_2}$
  operations indicates how the classical variable $x$ is used to
  control the measurement basis.  A value of $x = 0$ means that
  $HZ_{\alpha}$ is applied, while a value of $x = 1$ means that
  $HZ_{-\alpha}$ is applied.  Note that in the case of
  $HZ_{\alpha_1}$, we always have $x=0$, and so $HZ_{\alpha_1}$ is
  applied, as expected.
\label{fig:sq-csc-clas-control}}
\end{figure}

\begin{figure}%[ht]
\epsfig{file=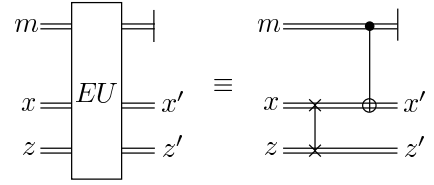} \caption{The classical error
update circuit, following Equations~(\ref{eq:EU-1}) and~(\ref{eq:EU-2}).
\label{fig:clas-err-update}}
\end{figure}

\begin{figure} %[ht]
\epsfig{file=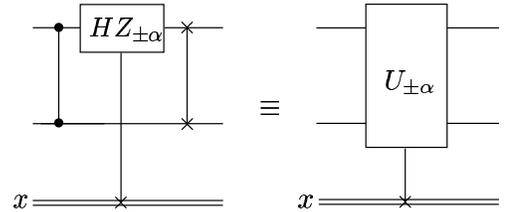} \caption{A definition to make other
circuits more compact. \label{fig:def-u}}
\end{figure}

\begin{figure}%[hb]
\scalebox{0.8}{ \epsfig{file=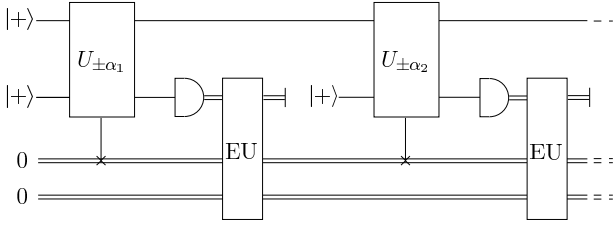} } \caption{The output
of Figure~\ref{fig:sq-csc-onebuff} is the same as the output of this
imperfect circuit. \label{fig:sq-csc-intermed}}
\end{figure}

The two classical variables $x$ and $z$ have been introduced into the
circuit in Figure~\ref{fig:sq-csc-clas-control}. The $E$rror $U$pdate
operation $EU$ updates $x$ and $z$ using Equations~(\ref{eq:EU-1})
and~(\ref{eq:EU-2}), and the variable $x$ is used to control the
rotations $H Z_{\pm \alpha_j}$. An explicit definition of $EU$ is
shown in Figure~\ref{fig:clas-err-update}.  The circuit of
Figure~\ref{fig:sq-csc-clas-control} can be made more compact by
introducing the notation of Figure~\ref{fig:def-u} for the classical
feedforward.  The resulting circuit is shown in
Figure~\ref{fig:sq-csc-intermed}.

% --------------------------------------------------------------------- %
% FIGURE: First appearance of classical control with the EU op          %
% --------------------------------------------------------------------- %

% --------------------------------------------------------------------- %
% FIGURE: Definition of the Error Update circuit                        %
% --------------------------------------------------------------------- %

% --------------------------------------------------------------------- %
% FIGURE: Definition of the U_\alpha operation                          %
% --------------------------------------------------------------------- %

We have made several modifications to the literal circuit
$\mathcal{L}$, but the output of the imperfect circuit shown in
Figure~\ref{fig:sq-csc-intermed} is still equivalent to the output of
the noisy literal circuit in Figure~\ref{fig:sq-csc-onebuff}. To
complete the construction of the blocks $B_{\alpha_j}$ referred to
earlier in Step~$3$, we introduce some additional circuit identities
whose effect is to compensate for the corrections $X^x Z^z$.  This
will enable us to make exact the correspondence with the gates $H
Z_{\alpha_j}$ in the quantum circuit $\mathcal{F_Q}$. 

% --------------------------------------------------------------------- %
% FIGURE: Intermediate Summary                                          %
% --------------------------------------------------------------------- %

\begin{figure} %[ht]
\scalebox{0.85}{ \epsfig{file=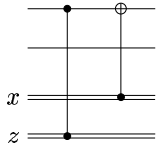} } \caption{The output
of Figure~\ref{fig:sq-csc-intermed} is unchanged if these perfect gates are
prepended at the beginning of the computation. \label{fig:prepend}}
\end{figure}

%
% at the beginning
%
First, at the beginning of the first block in
Figure~\ref{fig:sq-csc-intermed}, we prepend the gates illustrated in
Figure~\ref{fig:prepend}.  These are perfect gates, and can be
prepended without changing the output of
Figure~\ref{fig:sq-csc-intermed}, since the initial values of $x$ and
$z$ are both zero.

% --------------------------------------------------------------------- %
% FIGURE: Prepend for blocks                                            %
% --------------------------------------------------------------------- %

\begin{figure}%[hb]
\epsfig{file=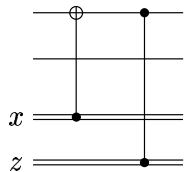} \caption{The effective output of
  Figure~\ref{fig:sq-csc-intermed} is unchanged if
  these perfect gates are appended at the end of the computation.
\label{fig:append}}
\end{figure}

Second, we modify the very final block in
Figure~\ref{fig:sq-csc-intermed} (which is not explicitly shown),
appending operations inverse to those in Figure~\ref{fig:prepend}, as
illustrated in Figure~\ref{fig:append}.  The reason we may do this is
as follows. If all the operations in the cluster-state computation are
implemented perfectly, then at the end of the computation the qubit
would be in the state $X^x Z^z|\psi\rangle$, where $|\psi\rangle$ is
the output from the corresponding perfect quantum circuit computation.
If we then measure this state in the computational basis, we can
compensate for the error operator $X^x Z^z$ by appropriate
post-processing of the measurement result, i.e., by adding $x$ to the
outcome of the measurement, modulo two.  This process of compensating
the measurement results is, however, equivalent to appending the
perfect gates illustrated in Figure~\ref{fig:append}, and dropping the
process of compensation.

% --------------------------------------------------------------------- %
% FIGURE: Append for blocks                                             %
% --------------------------------------------------------------------- %

\begin{figure} % [ht]
\epsfig{file=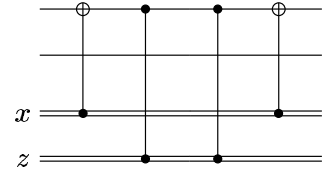} \caption{We insert these perfect gates
between each pair of blocks in Figure~\ref{fig:sq-csc-intermed}.
\label{fig:insert}}
\end{figure}

Our third and final modification is to insert the perfect gates of
Figure~\ref{fig:insert} between each pair of the repeating blocks in
Figure~\ref{fig:sq-csc-intermed}.  Another way of stating this is that
we insert the circuit of Figure~\ref{fig:append} at the \emph{end} of
every block in the computation, except the last block, where it has
already been inserted, and insert the circuit of
Figure~\ref{fig:prepend} at the \emph{beginning} of every block in the
computation, except the first, where it has already been inserted.
This insertion does not modify the output of the circuit, since, as is
apparent from Figure~\ref{fig:insert}, these gates cancel one another
out.

% --------------------------------------------------------------------- %
% FIGURE: Insertion between blocks                                      %
% --------------------------------------------------------------------- %

With these modifications, we see that the output of the noisy
cluster-state computation in Figure~\ref{fig:sq-csc-onebuff} is equivalent
to the output of the imperfect circuit illustrated in
Figure~\ref{fig:sq-csc-b}, where the operation $B_\alpha$ is defined in
Figure~\ref{fig:def-b}.

% --------------------------------------------------------------------- %
% FIGURE: Literal circuit expressed using the B_\alpha                  %
% --------------------------------------------------------------------- %

% --------------------------------------------------------------------- %
% FIGURE: Definition of the Block B_\alpha                              %
% --------------------------------------------------------------------- %

\begin{figure} %[ht]
\scalebox{0.9}{\epsfig{file=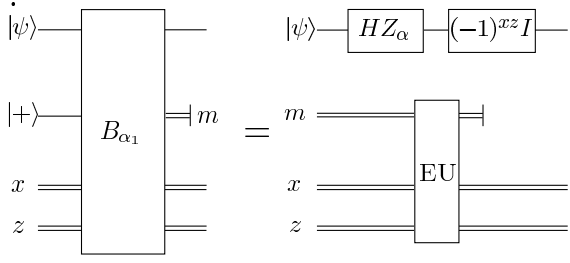}}

\caption{The output of the circuit on the left is identical to the output
  of the circuit on the right. For this identity we assume that all
  operations in both circuits are done perfectly --- there is no
  noise. \label{fig:b-equiv-hz}}
\end{figure}

%
% state the reason we've recast things in this way?
%
This completes the construction of the repeating blocks $B_\alpha$.
The circuit shown in Figure~\ref{fig:sq-csc-b} allows us a fiction of
identifying the topmost qubit $Q$ as a persistent data qubit carrying
the information in a cluster-state computation. We will see that
$B_\alpha$ effectively performs single-qubit gates on $Q$.  To see how
each $B_\alpha$ corresponds to a gate in the quantum circuit
$\mathcal{F_Q}$ consider the circuit identity shown in
Figure~\ref{fig:b-equiv-hz}. This identity shows that a \emph{perfect}
implementation of $B_\alpha$ is equivalent (up to a known global phase
factor) to the effect of applying a \emph{perfect} $HZ_\alpha$ gate to
the first qubit, initially in the state $|\psi\rangle$.  Intuitively,
then, we would expect that the result of the actual imperfections in
$B_\alpha$ would be to effect an imperfect $HZ_\alpha$ gate.  That is,
we obtain a way of translating our noisy cluster-state computation
into an equivalent noisy quantum circuit computation.

% --------------------------------------------------------------------- %
% FIGURE: Equivalence of B_\alpha to H Z_\alpha (with correction)       %
% --------------------------------------------------------------------- %

We have introduced the identity of Figure~\ref{fig:b-equiv-hz} for
motivational purposes only; we omit a proof, as we prove a stronger
result later.  Interested readers may wish to confirm this identity by
hand.  What we do now is find a way of showing quantitatively that
noise in the imperfect operation $B_\alpha$ may be mapped to noise in
the quantum circuit $\mathcal{F_Q}$.

The next step in our proof, Step~$4$, is to replace the classical
elements in $B_{\alpha_j}$ with unitary quantum equivalents to obtain
fully unitary blocks, which we denote $Q B_{\alpha_j}$. The reason for
doing so will become clear in the final step of the proof, where we
use the first unitary extension theorem, Theorem~\ref{thm:uet1}, to
compare an imperfect $Q B_{\alpha_j}$ to a noisy $H Z_{\alpha_j}$.

\begin{figure} %[ht]
\epsfig{file=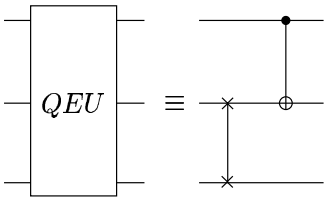} \caption{The definition of the quantum
error update operation, $QEU$. \label{fig:def-qeu}}
\end{figure}

The classical elements of $B_{\alpha_j}$ are the classical variables
$x,z$; the error update operation $EU$; and the classical controlled
$U_{\pm \alpha}$. These have all been replaced by quantum equivalents
to define $Q B_{\alpha_j}$ in Figure~\ref{fig:def-qb}. The quantum
error update operation $QEU$ is defined in Figure~\ref{fig:def-qeu},
and $U_{\pm \alpha}$ is defined as before but with a quantum control.
The bits carrying $x$ and $z$ have been replaced by qubits which we
label $X$ and $Z$.  We therefore assume that operations performed
on these qubits are noiseless, including the operations in $QEU$.

The output of the noisy literal circuit $\mathcal{L}$ is thus
equivalent to the output from the imperfect circuit in block unitary
form shown in Figure~\ref{fig:sq-csc-qb}.  The qubit labeled $Q$ is
persistent, and we can see that the cluster-state computation can be
thought of as the successive application of unitaries $Q
B_{\alpha_j}$.  These unitaries require three ancilla to perform an
operation on $Q$, but the following proposition shows that, when all
operations are done perfectly, the effect of each $Q B_{\alpha_j}$ on
$Q$ is identical to that of $H Z_{\alpha_j}$. 

% --------------------------------------------------------------------- %
% FIGURE: Definition of QB_\alpha                                       %
% --------------------------------------------------------------------- %

% --------------------------------------------------------------------- %
% FIGURE: Definition of QEU                                             %
% --------------------------------------------------------------------- %

% --------------------------------------------------------------------- %
% FIGURE: Literal circuit with QB_\alpha                                %
% --------------------------------------------------------------------- %

\begin{figure}[ht]
\scalebox{0.8}{ \epsfig{file=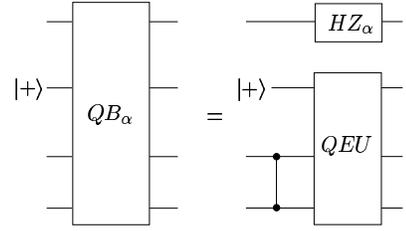}} \caption{The
circuit identity of  Proposition~\ref{prop:quantum-circuit-identity}. This
is a \emph{perfect} circuit identity, i.e., all elements in both circuits
are assumed to be performed without any noise. \label{fig:qb-equiv-hz}}
\end{figure}

\begin{proposition} {} \label{prop:quantum-circuit-identity}
  The circuit identity of Figure~\ref{fig:qb-equiv-hz} holds, where
  both circuits are assumed to be perfect.  All inputs are assumed to
  be arbitrary, except the fixed $|+\rangle$ input, as shown.
\end{proposition}

The proof of the circuit identity of Figure~\ref{fig:qb-equiv-hz} is
straightforward, but somewhat technical.  The details are sketched in
Appendix~\ref{app:single-qubit-identity}.

% --------------------------------------------------------------------- %
% FIGURE: Equivalence of QB_\alpha to H Z_\alpha                        %
% --------------------------------------------------------------------- %

The final step in establishing the noise correspondence for
single-qubit computation is to use
Proposition~\ref{prop:quantum-circuit-identity} and the first unitary
extension theorem, Theorem~\ref{thm:uet1}, to argue that an imperfect
implementation of $QB_{\alpha_j}$ is equivalent to a noisy operation
$H Z_{\alpha_j}$. Following the noise model of
Subsection~\ref{subsec:noise-model}, let $E$ be the environment
responsible for the noisy operations in the imperfect implementation
of $Q B_{\alpha_j}$. We denote this imperfect operation by $Q
B_{\alpha_j}^\prime$, a unitary acting on $QM_j XZ E$. By assuming
that the same environment $E$ is reused by all the imperfect
operations $QB_{\alpha_j}'$ we make the most pessimistic assumption we
can possibly make about noise in the cluster-state computation.  In
the more general multi-qubit situation this assumption will correspond
to assuming that all the qubits in the same level of the cluster share
the same environment.

Now, suppose $\eta$ is the maximal error strength in any of the noisy
operations making up the imperfect $QB_{\alpha_j}$.  That is, $\eta$
quantifies the strength of the noise in the cluster-state computation. Let
$c$ be the number of noisy operations in the imperfect $QB_{\alpha_j}$.
Then the chaining property, Proposition~\ref{prop:chaining}, implies that
\begin{eqnarray}
  \Delta_{QM_j XZ:E}(QB_{\alpha_j},QB_{\alpha_j}') \leq c \eta,
\end{eqnarray}
where $QB_{\alpha_j}$ is the \emph{perfect} $QB_{\alpha_j}$ operation.
Note our convention that in algebraic expressions we always use
$QB_{\alpha_j}'$ to refer to the imperfect operation, while in the
text we sometimes use $QB_{\alpha_j}$ to refer to the imperfect
operation, provided the context is clear.  It follows that there
exists a unitary operation $U_E$ on $E$ such that
\begin{eqnarray}
  \|QB_{\alpha_j}' - QB_{\alpha_j} \otimes U_E \| \leq c \eta.
\end{eqnarray}
(Note that $U_E$ may depend on $j$, but that dependence is not
important in the argument that follows, and so is suppressed.)

\begin{figure}[ht]
  \epsfig{file=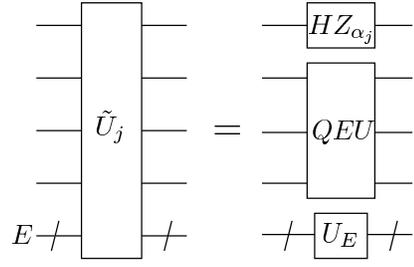} \caption{The definition of
    the operation $\tilde U_j$ The environment $E$ is shown at the
    bottom, where the wire with a slash through it indicates an
    arbitrary quantum system.
\label{fig:def-u-tilde}}
\end{figure}

So, if $Q B_{\alpha_j}^\prime$ is the noisy implementation of $Q
B_{\alpha_j}$, can we show that this corresponds to a noisy
implementation of $H Z_{\alpha_j}$? In Figure~\ref{fig:sq-csc-qb} we
see that the qubit $M_j$ is always initially prepared in the state
$|+\rangle$. This is an \emph{exact} statement, since imperfect
preparation of the state $|+\rangle$ is modeled as a perfect
preparation, followed by a noisy quantum memory step that is absorbed
into the imperfect operation $QB_{\alpha_j}$.  Define $S_j$ to be the
subspace of $QM_j XZ E$ in which $M_j$ is in the state $|+\rangle$,
and the other systems may be in an arbitrary state.  Then define $U_j
\equiv QB_{\alpha_j} \otimes U_E$ and $V_j \equiv Q
B_{\alpha_j}^\prime$. Define $\tilde{U}_j$ as shown in
Figure~\ref{fig:def-u-tilde}. By
Proposition~\ref{prop:quantum-circuit-identity} we see that $U_j
\big|_{S_j} = \tilde{U}_j \big|_{S_j}$ and so we can apply the first
unitary extension theorem, Theorem~\ref{thm:uet1}, to conclude that
there exists a unitary extension $\tilde V_j$ of $V_j|_{S_j}$ such
that
\begin{eqnarray}
  \| \tilde V_j - \tilde U_j\| & \leq & \| V_j - U_j \| \\
  & = &\|QB_{\alpha_j}' - QB_{\alpha_j} \otimes U_E \| \\
  & \leq & c \eta.
\end{eqnarray}
Applying Proposition~\ref{prop:error-reduction} we see that
\begin{eqnarray} \label{eq:noise-relation}
  \Delta_{Q:M_j XZ E}(HZ_{\alpha_j},\tilde V_j) \leq c \eta.
\end{eqnarray}

Next, define $M = M_1 \otimes M_2 \otimes \ldots$ to be the
combination of all the systems $M_j$.  Applying
Proposition~\ref{prop:error-ommission} and
Equation~(\ref{eq:noise-relation}) we see that
\begin{eqnarray} \label{eq:noise-relation-2}
  \Delta_{Q:E'}(HZ_{\alpha_j},\tilde V_j) \leq c \eta,
\end{eqnarray}
where $E' \equiv M X Y E$ is the \emph{effective environment} for the
qubit $Q$, and we have extended $\tilde V_j$ to act in the natural way
on $E'$, i.e., $\tilde V_j$ acts trivially on systems $M_k$ such that
$k \neq j$.  Because $\tilde V_j|_{S_j} = V_j|_{S_j}$ we see that the
noisy implementation $V_j = QB'_{\alpha_j}$ of $QB_{\alpha_j}$ is
exactly equivalent to a noisy implementation $\tilde V_j$ of $H
Z_{\alpha_j}$ of strength at most $c \eta$.

% --------------------------------------------------------------------- %
% FIGURE: Definition of extension \tilde{U}                             %
% --------------------------------------------------------------------- %

\begin{figure} [ht]
\epsfig{file=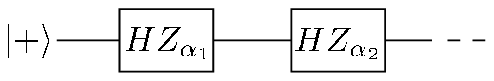} 
\caption{This noisy circuit is
  equivalent to the noisy one-buffered cluster-state computation of
  Figure~\ref{fig:sq-csc-onebuff}.
\label{fig:sq-csc-final}}
\end{figure}

To conclude, we have shown that if the one-buffered cluster-state
computation depicted in Figure~\ref{fig:sq-csc-onebuff} is implemented
with noise of strength at most $\eta$ in each operation, then the
output of that computation is equivalent to the output of the noisy
quantum circuit in Figure~\ref{fig:sq-csc-final}, where each operation
is performed with noise of strength at most $c \eta$.  As we have
described it, $c$ is the number of noisy operations in the imperfect
operation $QB_\alpha$, i.e., $c \approx 10^1$.  In actual
implementations it would be possible to directly evaluate the total
strength of the noise in the imperfect operation $QB_{\alpha}$,
resulting in a more accurate (and better, from the point of view of
the threshold) value for $c$.

A interesting feature of our argument is that even if the various
physical operations involve only Markovian noise, the corresponding
effective noise in the implementation of $H Z_{\alpha_j}$ is
inherently non-Markovian.  The reason is that the qubits $X$ and $Z$
associated with the classical variables $x$ and $z$ are part of the
effective environment of $Q$ at every stage of the computation, due to
the necessity of feeding forward the measurement results.  This is the
reason we need to use the non-Markovian threshold result of Terhal and
Burkard~\cite{Terhal04a}.

%% --------------------------------------------------------------------- %
%% FIGURE: Stuff                                                         %
%\begin{figure} %[ht]
%\epsfig{file=./figures/sq-csc-qeu.eps} \caption{The state of $Q$ at the
%end of this circuit is the same as the
%  state of $Q$ at the end of the imperfect circuit in
%  Figure~\ref{fig:sq-csc-qb}.
%\label{fig:sq-csc-qeu}}
%\end{figure}
%% --------------------------------------------------------------------- %

% --------------------------------------------------------------------- %
% FIGURE: Noisy single qubit quantum circuit                            %
% --------------------------------------------------------------------- %

% \end{document}

% %%%%%%%%%%%%%%%%%%%%%%%%%%%%%%%%%%%%%%%%%%%%%%%%%%%%%%%%%%%%%%%%%%%
% %%%%%%%%%%%%%%%%%%%%%%%%%%%%%%%%%%%%%%%%%%%%%%%%%%%%%%%%%%%%%%%%%%%
% %%%%%%%%%%%%%%%%%%%%%%%%%%%%%%%%%%%%%%%%%%%%%%%%%%%%%%%%%%%%%%%%%%%
% %%%%%%%%%%%%%%%%%%%%%%%%%%%%%%%%%%%%%%%%%%%%%%%%%%%%%%%%%%%%%%%%%%%
% %%%%%%%%%%%%%%%%%%%%%%%%%%%%%%%%%%%%%%%%%%%%%%%%%%%%%%%%%%%%%%%%%%%
% %%%%%%%%%%%%%%%%%%%%%%%%%%%%%%%%%%%%%%%%%%%%%%%%%%%%%%%%%%%%%%%%%%%
% %%%%%%%%%%%%%%%%%%%%%%%%%%%%%%%%%%%%%%%%%%%%%%%%%%%%%%%%%%%%%%%%%%%
% %%%%%%%%%%%%%%%%%%%%%%%%%%%%%%%%%%%%%%%%%%%%%%%%%%%%%%%%%%%%%%%%%%%

% -------------------------------------------------------------------- %
% MULTI-QUBIT NOISE CORRESPONDENCE
% -------------------------------------------------------------------- %

\subsection{Noise correspondence for multi-qubit computations}
\label{subsec:multi-qubit-step}

\begin{figure}[ht]
\epsfig{file=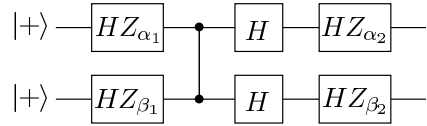}

\caption{An illustrative two-qubit quantum circuit $\mathcal{F_Q}$ using
canonical fault-tolerant gates. \label{fig:multi-qubit-circuit}}
\end{figure}

\begin{figure}[ht]
\epsfig{file=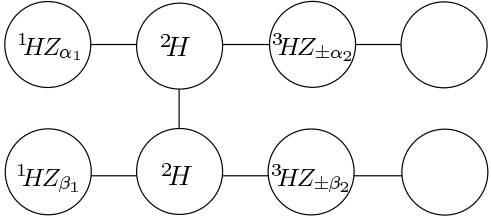} \caption{A noisy two-qubit
cluster-state computation. \label{fig:mq-csc}}
\end{figure}

\begin{figure}[h]
  \scalebox{0.95}{\epsfig{file=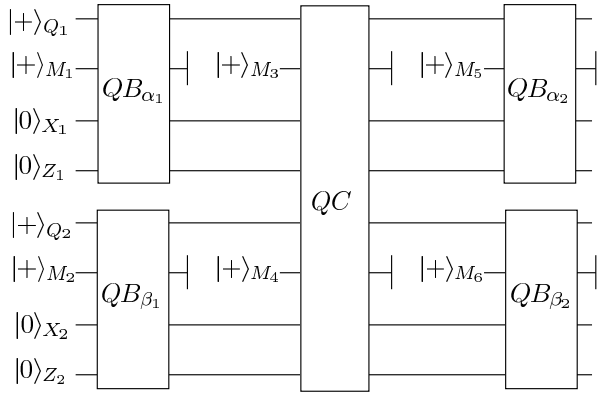}}
  \caption{The output of this imperfect circuit is equivalent to the
    output of the noisy two-qubit cluster-state computation in
    Figure~\ref{fig:mq-csc}.  The operations $QB_{\alpha_j},
    QB_{\beta_j}$ are as defined in the previous subsection, while
    $QC$ is defined in Figure~\ref{fig:def-qc}.
\label{fig:mq-csc-onebuff}}
\end{figure}

%
% what we do
%
In this subsection we extend the ideas of the previous subsection,
explaining how noise in a multi-qubit one-buffered cluster-state
quantum computation may be mapped to equivalent noise in the
corresponding quantum circuit.  The ideas used to do this are the same
as were used in the proof for single-qubit cluster-state computations.
As a result, we merely sketch out how the proof goes for a specific
example, with the general proof following similar lines.

\begin{figure}[ht]
\scalebox{0.85}{\epsfig{file=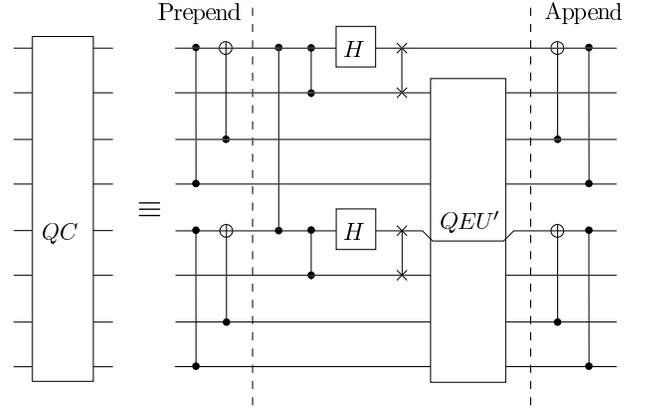}}

\caption{The definition of the imperfect operation $QC$ appearing in
  Figure~\ref{fig:mq-csc-onebuff}. We have prepended and appended the
  same fictitious elements used in the single-qubit case in
  Figures~\ref{fig:prepend} and~\ref{fig:append}. The gate
  $QEU^\prime$ is the two-qubit extension of the error update, and is
  shown in detail in Figure~\ref{fig:qeu-prime-def}.  The only noisy
  operations are the Hadamard gates, and the {\sc cphase} gate between
  the first and the fifth qubits.
\label{fig:def-qc}}
\end{figure}

%
% introduce our example
%
The noise correspondence is established via the same process followed
in Subsection~\ref{subsec:single-qubit-step}. We begin with a
multi-qubit quantum circuit like the circuit $\mathcal{F_Q}$ shown in
Figure~\ref{fig:multi-qubit-circuit}. The cluster-state computation
$\mathcal{C}$ simulating this circuit is shown in
Figure~\ref{fig:mq-csc}. As before, we construct a literal circuit and
rearrange it into the block form where each block is identifiable with
a gate in the quantum circuit $\mathcal{F_Q}$. We have omitted the
details of this rearrangement, and presented the final block form
equivalent to the literal circuit in Figure~\ref{fig:mq-csc-onebuff}.
The blocks $QB_{\alpha_j}, QB_{\beta_j}$ correspond to the
single-qubit gates $H Z_{\alpha_j}, H Z_{\beta_j}$ in $\mathcal{F_Q}$,
as in the previous subsection, and the effects of noise in these
blocks has already been considered. The new element is the unitary
block $QC$, which is shown in detail in Figure~\ref{fig:def-qc}. The
following proposition shows that a perfect implementation of $QC$
effects the operation $(H \otimes H)${\sc cphase} on the qubits $Q_1$
and $Q_2$.

%
% introduce the proposition
%

\begin{figure}[ht]
\epsfig{file=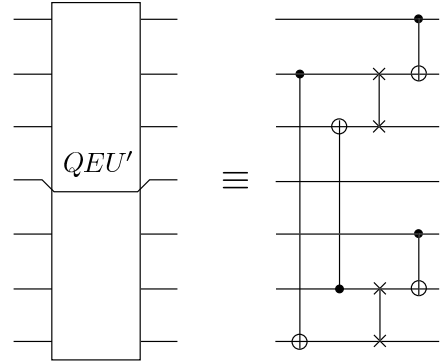} \caption{The definition of the
two-qubit quantum error update operation analogous to that of
Figure~\ref{fig:def-qeu}. \label{fig:qeu-prime-def}}
\end{figure}

\begin{figure}[h]
\epsfig{file=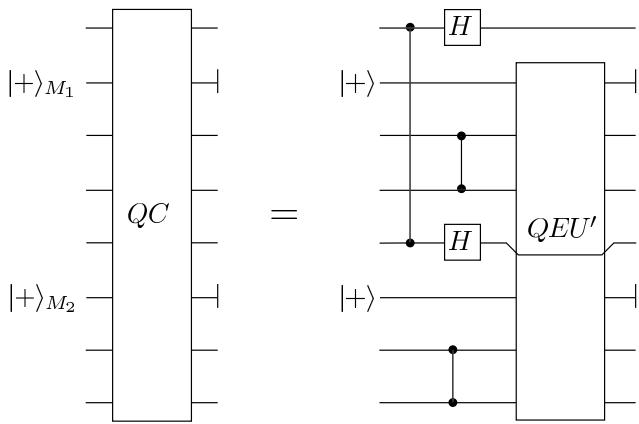} \caption{The circuit identity
of Proposition~\ref{prop:two-qubit-quantum-circuit-identity}. This is a
\emph{perfect} circuit identity, i.e., all elements in both circuits are
assumed to be performed without any noise. \label{fig:qc-equiv-circuit}}
\end{figure}

\begin{proposition} {} \label{prop:two-qubit-quantum-circuit-identity}
  The circuit identity of Figure~\ref{fig:qc-equiv-circuit} holds, where
  both circuits are assumed to be perfect.
\end{proposition}

\textbf{Proof:} The proof of this identity uses the same techniques as
the proof of Proposition~\ref{prop:quantum-circuit-identity}, and is
omitted.  The identity may easily be verified using any of the
standard computer algebra packages.

\textbf{QED}

This proposition plays a similar role to that played by
Proposition~\ref{prop:quantum-circuit-identity} in establishing the
single-qubit noise correspondence. In this case the noise
correspondence will again follow from the first unitary extension
theorem.

Following the noise model of Subsection~\ref{subsec:noise-model}, we
introduce environments $E_1$ and $E_2$ associated with the respective
levels in the cluster-state computation of Figure~\ref{fig:mq-csc}.
Under the locality assumption an imperfect $Q B_{\alpha_1}$, for
example, is thus represented as a unitary operator $Q
B_{\alpha_1}^\prime$ acting on $Q_1 M_1 X_1 Z_1 E_1$, which results in
an effective environment for the corresponding $H Z_{\alpha_1}^\prime$
of $\tilde E_1 = M_1 X_1 Z_1 E_1$. As in the single-qubit case, the
noise strength in the operations of the form $HZ_\alpha$ is at most $c
\eta$, where $c \approx 10^1$ is a small constant, and $\eta$ is the
noise strength in the operations used to implement the cluster-state
computation.

An imperfect $QC$ is similarly represented as a unitary $QC^\prime$
which involves \emph{both} the environments $E_1$ and $E_2$. Using the
first unitary extension theorem we can show that this results in an
effective environment $\tilde E_{12} = M_3 X_1 Z_1 M_4 X_2 Z_2 E_1
E_2$ for the operation $(H \otimes H)${\sc cphase} in $\mathcal{F_Q}$.
It follows that the noise strength in the $(H \otimes H)${\sc cphase}
gate is at most $c' \eta$, where $c'$ is the total number of noisy
operations in the imperfect operation $QC$, and again is of order
$10^1$.

%
% summing up
%
Summing up, suppose we perform a noisy one-buffered implementation of
a cluster-state computation, satisfying the noise model described in
Subsection~\ref{subsec:noise-model}, and with $\eta$ the maximal noise
strength in any operation.  Then we can show that this noisy
computation is equivalent to performing the corresponding quantum
circuit with noise satisfying the locality assumption, and of strength
at most $c'' \eta$, where $c'' \approx 10^1$ is some constant.  This
completes the proof that quantum circuits may be simulated in a
fault-tolerant fashion using a one-buffered cluster-state computation.

One remarkable feature of our proof is that it goes through unchanged
even if we allow noise to occur in the classical computations and
feedforward.  It is easy to see that such noise simply causes an
additional contribution to the strength of the noise in the
corresponding quantum circuit, and thus a decrease in the effective
threshold.  This feature of the proof also carries over to the
threshold theorem for optical cluster-state computation presented in
the next section.  Thus, our results show that not only can
cluster-state computation be made resilient against the effects of
unitary non-Markovian errors, it can even be made resilient against
the effects of noise in the classical parts of the computation.

%% file: ftqc-cs-ocs.tex
\section{Fault-tolerance with optical cluster states}
\label{sec:optical-ft}

In this section we explain how the ideas of
Section~\ref{sec:deterministic} can be extended to enable fault-tolerant
simulation of quantum circuits using \emph{optical} cluster states.
The main challenge in proving this result is the non-deterministic
nature of the entangling gates used in optical cluster-state
computation.  We show that this challenge can be met by using those
non-deterministic gates to add additional pieces to the cluster in a
near-deterministic fashion using what we call the \emph{dangling node}
implementation of optical cluster-state quantum computation.  On those
rare occasions when the addition to the cluster fails, one simply
accepts failure and moves on, regarding the failure as a small amount
of additional noise in a deterministic preparation of the cluster
state.  This allows us to apply the results of the
Section~\ref{sec:deterministic} to deduce a fault-tolerant threshold for
optical cluster states. 

Although we phrase our results in terms of optics, there may be other
natural contexts in which our results apply.  The crucial fact about
the optical implementation that we use is the existence of a
postselected non-deterministic {\sc cphase} gate which, when it fails,
effects a measurement in the computational basis.  The threshold
results we apply would apply equal well to other implementations which
share this feature.

We begin our account in Subsection~\ref{subsec:ocs-overview} with an
overview of the main elements of the proof.  In particular, we explain
the dangling node implementation in detail, and explain heuristically
how noise in the dangling node implementation maps to noise in the
quantum circuit model.  Subsection~\ref{subsec:ocs-noise-model}
discusses the noise model we use in our description of optical
cluster-state computation.  An important issue in this discussion is
noise due to photon loss.  As described in the introduction to this
paper, the noise model used by Terhal and Burkard~\cite{Terhal04a}
does not explicitly allow for leakage errors, and thus noise due to
photon loss is not directly addressed by our results.  Nonetheless, we
expect that with some straightforward modifications of the results
in~\cite{Terhal04a} it will be possible to cope with the problem of
photon loss and other forms of leakage error in optical cluster-state
computation.  Subsection~\ref{subsec:ocs-intermediate} proves a
rigorous threshold for what we call the \emph{two-at-a-time}
implementation of cluster-state computation.  The two-at-a-time
implementation makes use of deterministic {\sc cphase} gates, and thus
is not of immediate relevance for optical cluster-state computing.
Its interest arises from the fact that it is intermediate in
complexity between the one-buffered implementation studied in
Section~\ref{sec:deterministic}, and the full dangling node
implementation of optical cluster-state computation.  Analysing the
two-at-a-time implementation thus provides a useful stepping stone on
the way to understanding fault-tolerance in the dangling node
implementation.  Subsection~\ref{subsec:ocs-postselected} introduces
some useful nomenclature for describing postselected quantum gates,
and proves a simple lemma about such gates.  The threshold proof for
optical cluster-state computation is completed in
Subsection~\ref{subsec:ocs-threshold}, which explains how noise in the
dangling node implementation may mapped to equivalent noise in the
quantum circuit model.

\subsection{Overview}
\label{subsec:ocs-overview}

%
% basic idea
%
The broad structure of the threshold proof for optics is similar to
the proof for the deterministic one-buffered implementation discussed
in Section~\ref{sec:deterministic}.  We take a quantum circuit that we
wish to simulate, and turn it into a fault-tolerant quantum circuit,
using the standard prescriptions for fault-tolerance.  We then convert
the fault-tolerant circuit into an equivalent cluster-state
computation, and then specify in detail a specific implementation
protocol for performing that computation.  In this case that will be a
dangling node implementation.  We now describe how the dangling node
implementation works.  We will show in later subsections that noise in
the dangling node implementation can be mapped onto equivalent noise
in the original fault-tolerant quantum circuit.  This enables us to
deduce that the output of the noisy optical cluster-state computation
is equivalent to the output of a noisy fault-tolerant quantum circuit
computation, and thus to obtain a threshold result.

\begin{figure}
\epsfig{file=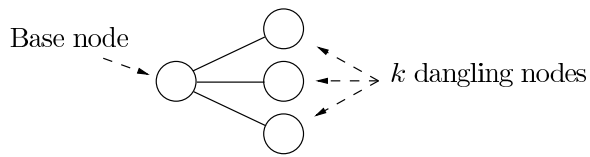} \caption{The microcluster
used in the dangling node
  implementation of cluster-state computation.  There are $k$ dangling
  nodes, in general; in this example, $k = 3$.
 \label{fig:microcluster-basic}}
\end{figure}

%
% single-qubit case
%
We begin by describing the dangling node implementation for the
simplest case of a single-qubit cluster-state computation.  The basic
idea is to to build the cluster up using small cluster states that we
call \emph{microclusters}\footnote{A similar microcluster construction
  was used for different purposes in~\cite{Nielsen04b}.}.  The basic
microcluster is illustrated in Figure~\ref{fig:microcluster-basic}.  It
includes a single \emph{base node}, on the left, and $k$
\emph{dangling nodes}, on the right, where $k$ is some fixed constant.

%
% preparation of microclusters
%
Optically, we may prepare microclusters by sequentially performing $k$
non-deterministic {\sc cphase} gates.  These may be either the
non-deterministic KLM {\sc cphase} gate, or one of its more efficient
(but still non-deterministic) modern descendants.  Since $k$ is a
constant, the expected number of operations required to form a
microcluster is also constant.  In practice, there are likely to be
much more efficient means of preparing a microcluster than this
procedure.  However, for the purposes of this paper we shall not be
concerned with optimizing the formation of the microcluster.

%
% how things actually work
%
The dangling node implementation of a single-qubit cluster-state
quantum computation works as follows.  The first step is to prepare a
single microcluster, as illustrated in
Figure~\ref{fig:microcluster-basic}.  This forms the basis from which a
larger cluster state will gradually be grown by adjoining extra
microclusters.

\begin{figure}
\epsfig{file=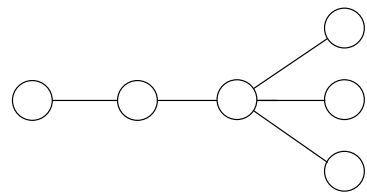}
\caption{The cluster state which results after successfully adjoining
  a microcluster to the initial microcluster in
  Figure~\ref{fig:microcluster-basic}, deleting the extra nodes, and
  applying appropriate local operations.
 \label{fig:larger-cluster-success}}
\end{figure}

\begin{figure}
\epsfig{file=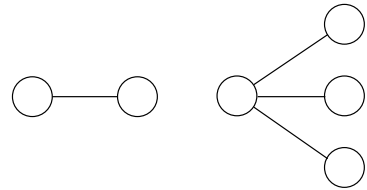}
\caption{The cluster state which results after failing $k-1$ times to adjoin
  a microcluster to the initial microcluster in
  Figure~\ref{fig:microcluster-basic}.
 \label{fig:larger-cluster-failure}}
\end{figure}

%
% second step
%
The second step of the dangling node implementation is to attempt to
adjoin microclusters to each of the first $k-1$ dangling nodes of the
first microcluster (i.e., all but the last of the dangling nodes),
using non-deterministic {\sc cphase} gates.  If one of these attempts
to adjoin should succeed, then we stop, and use computational basis
measurements and single-qubit operations to ensure that we end up with
the larger cluster state shown in
Figure~\ref{fig:larger-cluster-success}.  To ensure that this works,
it is critical that we use one of the special KLM non-deterministic
{\sc cphase} gates, which, as noted in
Section~\ref{sec:cluster-state-qc}, have the property that failure of
the gate results in a computational basis measurement, and removal of
a qubit from the cluster.  If all $k-1$ attempts to adjoin fail, then
we simply accept the failure, and apply the appropriate single-qubit
operation to end up with the larger cluster state shown in
Figure~\ref{fig:larger-cluster-failure}.  Note that if the
non-deterministic KLM {\sc cphase} gate fails with probability $p_f$,
then the probability of successfully adjoining a microcluster, as in
Figure~\ref{fig:larger-cluster-success}, is $1-p_f^{k-1}$, while the
probability of failing, as in Figure~\ref{fig:larger-cluster-failure},
is $p_f^{k-1}$.  When the adjoinment succeeds, we effectively add two
perfect extra layers of qubits to the cluster.  Failure introduces a
defect into the cluster, but we can ensure this occurs with small
probability by choosing $k$ to be large.

\begin{figure}
\epsfig{file=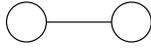}
\caption{The simpler two-qubit microcluster adjoined to the cluster state
  at the very \emph{end} of a single-qubit cluster-state computation,
  in place of the microcluster in Figure~\ref{fig:microcluster-basic}.
 \label{fig:simpler-microcluster}}
\end{figure}

%
% third step
%
The third step of the dangling node implementation is to perform the
first two layers of single-qubit measurements in the cluster-state
computation.  We then go back to the second step, effectively adding
another two layers of qubits into the cluster, and so on, through the
entire course of the computation.  The only variation comes at the
very end of the computation, where there is no need to adjoin a
microcluster of the form in Figure~\ref{fig:microcluster-basic}, but
instead we can add the simpler two-qubit microcluster illustrated in
Figure~\ref{fig:simpler-microcluster}.

\begin{figure}
\epsfig{file=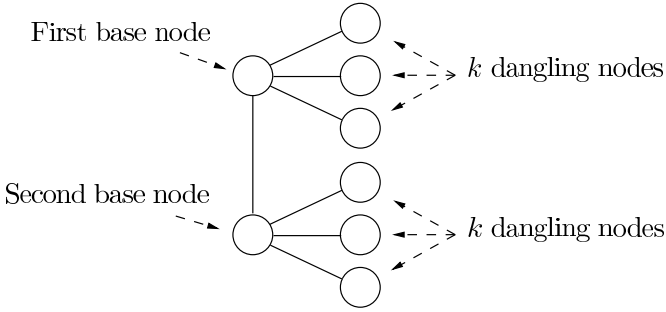}
  \caption{This microcluster is used in the dangling node
    implementation of multi-qubit cluster-state quantum computations.
 \label{fig:other-microcluster}}
\end{figure}

%
% segue into mult-qubit cluster
%
We have seen how a single-qubit cluster-state computation may be
performed in the dangling node implementation; what about multi-qubit
cluster-state computations?  The key is to introduce a third type of
microcluster, illustrated in Figure~\ref{fig:other-microcluster}.  Once
again, note that these microclusters may be prepared in constant
expected time using non-deterministic {\sc cphase} gates.

\begin{figure}
\epsfig{file=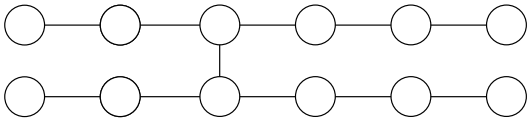}
\caption{An example cluster-state computation.  We have omitted explicit
  description of the measurements performed, since it is the formation
  of the cluster that we wish to concentrate on, rather than the
  details of the measurements.
 \label{fig:simple-example-ocs}}
\end{figure}

%
% example: first step
%
Consider the cluster-state computation depicted in
Figure~\ref{fig:simple-example-ocs}.  We may implement this computation
as follows.  The first step is to prepare the first two layers of the
cluster.  This may be done using the non-deterministic {\sc cphase}
gate in constant time, as described earlier.

\begin{figure}
\epsfig{file=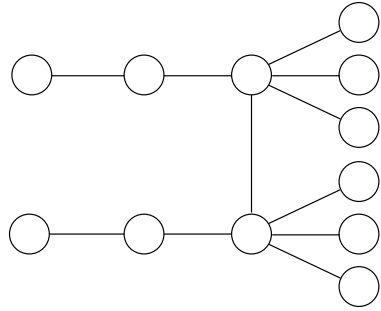}
\caption{The cluster after successfully adding the microcluster of
Figure~\ref{fig:other-microcluster}.
 \label{fig:simple-example-ocs-intermediate-success}}
\end{figure}

\begin{figure}
\epsfig{file=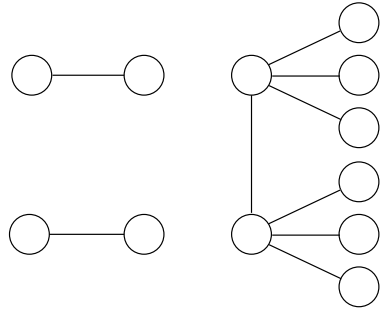}
\caption{The cluster after failing to add the microcluster of
Figure~\ref{fig:other-microcluster}.
 \label{fig:simple-example-ocs-intermediate-failure}}
\end{figure}

%
% second step
%
The second step of the dangling node implementation is to add in two
additional layers to the cluster.  This is done in a similar fashion
to the adjoinment of a microcluster in the single-qubit case.  In this
case, success requires that we adjoin the microcluster in
Figure~\ref{fig:other-microcluster} to two different dangling nodes, one
on the top level of the computation in
Figure~\ref{fig:simple-example-ocs}, the other to the bottom level.
Success in adjoinment at both levels results in the cluster of
Figure~\ref{fig:simple-example-ocs-intermediate-success} being prepared.
A simple calculation shows that this occurs with probability $1-2
p_f^{k-1}+p_f^{2k-2}$, which approaches $1$ rapidly as $k$ becomes
large.  If, on the other hand, either adjoinment should fail, we simply
declare failure overall, resulting in the cluster of
Figure~\ref{fig:simple-example-ocs-intermediate-failure} being prepared.
Note that if one part of the microcluster is successfully adjoined,
but the other part fails, then it may be necessary to delete some
nodes of the cluster using computational basis measurements and local
$Z$ operations in order to obtain the cluster of
Figure~\ref{fig:simple-example-ocs-intermediate-failure}.

%
% third and fourth steps
%
The third step in the dangling node implementation of the computation
in Figure~\ref{fig:simple-example-ocs} is to perform the first two
layers of single-qubit measurements.  The fourth step is to add in the
final two layers of the cluster, which is done using the same
procedure as for single-qubit computations.  The fifth and final step
of the implementation is to do the final four layers of single-qubit
measurements in the standard way.

%
% generalization
%
The generalization of the dangling node implementation to arbitrary
multi-qubit cluster-state computations follows the same lines,
alternating attempts to adjoin microclusters with two layers of
single-qubit measurements.  Each successful adjoinment of a
microcluster adds two perfect extra layers to one or more levels of
the cluster, while failure to adjoin correctly introduces a defect
into the cluster, but occurs with probability $2
p_f^{k-1}-p_f^{2k-2}$.

%
% heuristic noise mapping
%
How does noise in the dangling node implementation map to noise in the
quantum circuit model?  A heuristic argument is as follows.  Roughly
speaking, each gate in the original circuit is simulated by adjoining
a microcluster and performing some single-qubit measurements.
Adjoining a microcluster involves up to $c_1 k^2$ physical operations,
where $c_1$ is some positive constant in the range $10^0$-$10^2$.  If
each operation is performed with error strength $\eta$, then the total
associated noise is at most $c_1 k^2 \eta$.  There is also an
intrinsic failure probability, due to the non-determinism of the
gates, which causes defects in the cluster.  This probability is
bounded by $2 p_f^{k-1}$.  Finally, the noise contribution due to the
single-qubit measurements scales as $c_2 \eta$, where $c_2$ is a
constant of order $10^0$.  In consequence we expect that noise of
strength $\eta$ in the dangling node implementation will map to
equivalent noise of strength $c_1 k^2 \eta + c_2 \eta+ 2p_f^{k-1}$ in
the quantum circuit model.  It follows that if $\eta_{\rm th}$ is the
threshold in the quantum circuit model then provided $\eta$ satisfies
$c_1 k^2 \eta + c_2 \eta+ 2p_f^{k-1} \leq \eta_{\rm th}$
fault-tolerant computation is possible using optical cluster-state
computation.  The goal of the next four subsections is to make this
heuristic argument rigorous.  We will see that the rigorous
conclusions are in qualitative agreement with this heuristic analysis,
with some minor quantitative changes.

An issue we have glossed over in our discussion is that the dangling
node implementation restricts the structure of the clusters that may
be formed.  In particular, connections between levels of the cluster
can only be formed within odd numbered layers of the cluster.  This
has the effect of slightly restricting the quantum circuit operations
that may be directly simulated with such a cluster.  Using an argument
analogous to that in Subsection~\ref{subsec:circuit-threshold} one can
verify that this restricted canonical set of operations can be used to
prove a threshold theorem analogous to
Theorem~\ref{thm:threshold-canonical}.  We omit the details of this
argument, which is straightforward.

\subsection{The noise model in optical cluster-state computation}
\label{subsec:ocs-noise-model}

We assume that noise in optical cluster-state computation follows
essentially the same model as was introduced in
Subsection~\ref{subsec:noise-model} for cluster-state computation with
deterministic {\sc cphase} gates.  In particular, we assume that noisy
preparation and measurement can be modeled as perfect operations,
accompanied by noisy quantum memory steps.  To model noisy unitary
operations we assume that each level in the cluster has its own
environment, and that non-interacting levels have non-interacting
environments.  The only new element that needs to be accounted for is
when unitary operations are performed which involve the ancilla used
during the non-deterministic {\sc cphase} gate.  To cope with this, we
make the pessimistic assumption that these ancilla share a common
environment with whichever levels of the cluster are involved in the
attempted {\sc cphase} gate.

This noise model omits a significant possible source of noise, that of
photon loss.  The basic problem, as alluded to earlier, is that the
noise model we are using, based on that in~\cite{Terhal04a}, does not
allow leakage errors.  Noise is assumed to arise from the interaction
of a qubit with some environment.  In reality, noise may sometimes
have a rather different nature.  In particular, a qubit is sometimes a
two-dimensional subspace of a larger physical state space, and noise
may be due not to interaction with an environment, but rather to
leakage of the state of the qubit into some other part of the state
space.  This is exactly the type of error caused by photon loss in
optical quantum computation.

How to deal with this type of leakage error is well-understood in the
theory of quantum error-correction, and is addressed by several of the
standard threshold theorems.  In particular, loss detection techniques
are used in the threshold analysis~\cite{Knill00a} accompanying the
original KLM proposal for optical quantum computation.  This type of
error is not, however, explicitly addressed by the threshold theorem
of~\cite{Terhal04a}.  Although we believe that the result
of~\cite{Terhal04a} can likely be adapted to cope with such leakage
errors, we have not yet performed a complete analysis.  Such an
analysis will appear in future work.  In the meantime, our results
apply to the more restricted model of noise without leakage.

\subsection{The two-at-a-time implementation of cluster-state quantum
computation}
\label{subsec:ocs-intermediate}

In Section~\ref{sec:deterministic} we explained how noise in a
one-buffered implementation of cluster-state computation may be mapped
to noise in a quantum circuit computation.  In this section we'll
extend those results to a more complex implementation of cluster-state
computation, which we call a two-at-a-time implementation.

\begin{figure}
  \epsfig{file=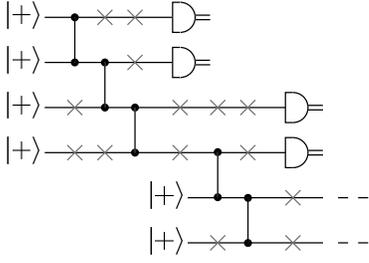} \caption{The literal circuit
    for a two-at-a-time implementation of a single-qubit cluster-state
    computation, with the noisy quantum memory steps explicitly
    indicated by crosses.  For simplicity we have combined the
    single-qubit rotations and computational basis measurements into a
    single-qubit measurement in some other basis, not explicitly
    specified.
 \label{fig:two-at-a-time}}
\end{figure}

A two-at-a-time implementation is similar to the one-buffered
implementation, except now qubits are added into the cluster two
layers at a time, and the single-qubit measurements are performed two
layers at a time.  In particular, we assume that deterministic {\sc
  cphase} gates are available in a two-at-a-time implementation.  More
explicitly:
\begin{itemize}
\item Prepare layers one through four of the cluster, using
  $|+\rangle$ preparations and (deterministic) {\sc cphase} gates.
\item Perform the first two layers of measurements.
\item Prepare layers five and six, using $|+\rangle$ preparations and
  {\sc cphase} gates to adjoin the extra layers.
\item Perform the third and fourth layers of measurements.
\item Keep alternating the preparation of two extra layers with the
  measurement of two extra layers, until the end of the computation.
\end{itemize}
The literal circuit for a two-at-a-time implementation of a
single-qubit cluster-state computation is shown in
Figure~\ref{fig:two-at-a-time}.  In this figure we have explicitly drawn
in the noisy quantum memory steps, using little crosses to indicate
when a quantum memory step is being performed.  Including these
explicitly makes it easier to map this implementation onto an
equivalent one-buffered implementation.

\begin{figure}
\scalebox{0.7}{ \epsfig{file=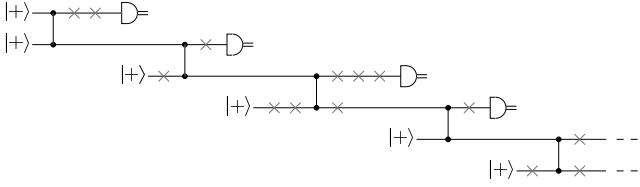}}
\caption{The output of this circuit is equivalent to
  the output of the literal circuit for the two-at-a-time implementation
  in Figure~\ref{fig:two-at-a-time}.
 \label{fig:one-buffered-equivalent}}
\end{figure}

Commuting the operations in Figure~\ref{fig:two-at-a-time} forward in
time, we see that that the output of Figure~\ref{fig:two-at-a-time} is
equivalent to the output of the circuit in
Figure~\ref{fig:one-buffered-equivalent}.  Note that doing this
commutation requires the use of
Proposition~\ref{prop:commute-forward}, since noisy operations on
different qubits that are on the same level of the cluster may
potentially involve the same environment, and so may not commute.
Fortunately, Proposition~\ref{prop:commute-forward} allows this
commutation to be performed without increasing the strength of the
underlying noise.

Inspection of Figure~\ref{fig:one-buffered-equivalent} reveals that this
circuit may be regarded as the literal circuit for a one-buffered
implementation of a cluster-state computation, and thus is equivalent
to a noisy quantum circuit, using the results of
Section~\ref{sec:deterministic}.  A similar argument can be used to map
noise in a multi-qubit two-at-a-time implementation of a cluster-state
computation to equivalent noise in a quantum circuit.  We omit the
details of the argument, which is a straightforward extension of the
single-qubit case.

\subsection{Postselected gates}
\label{subsec:ocs-postselected}

%
% what we do in this section
%
In order to map noise in optical cluster-state computation into the
quantum circuit model, we need a simple lemma about postselected
quantum gates.  In this subsection we provide a formal definition of
what we call a \emph{unitary postselected gate}, and prove the
required lemma.  The unitary postselected gates discussed here differ
in an important respect from the postselected gates in the discussion
of optical quantum computing in Section~\ref{sec:cluster-state-qc}.  In
the optical gates, success of the quantum gate is conditional on some
measurement outcome occurring.  In the present scenario the
postselected gate is all-unitary, i.e., no measurement is involved.
The connection between the two types of postselection is made by
replacing the measurement by an equivalent unitary process; we will
see an explicit example of how this works in the next subsection.

%
% definition of
%
Let $U$ be a unitary gate acting on two registers, labeled $A$ and
$B$.  $B$ is initially in some fixed state $|\beta\rangle$, while $A$
may be in an arbitrary state $|\psi\rangle$.  Let $V$ be a unitary
gate acting on register $A$ alone.  $U$ is said to be a \emph{unitary
  postselected gate that implements $V$ with probability $p$} if for
all $|\psi\rangle$,
\begin{eqnarray}
  U|\psi\rangle |\beta\rangle = \sqrt{p}V|\psi\rangle |\beta'\rangle +
  \sqrt{1-p} |\psi'\rangle |\beta''\rangle,
\end{eqnarray}
where $|\beta'\rangle$ is some fixed state, and $|\beta''\rangle$ is
orthonormal to $|\beta'\rangle$.

\begin{lemma} {} \label{lemma:unitary-approximation}
  Let $U$ be a unitary postselected gate implementing $V$ with
  probability $p$, on registers $A$ and $B$, and when $|\beta\rangle$
  is input to the register $B$.  Let $S$ be a subspace of the state
  space for register $A$, and let $T$ be the subspace of the total
  state space for $AB$ that contains states of the form $|\psi\rangle
  |\beta\rangle$, where $|\psi\rangle \in S$. Then there exists a
  unitary $W$ acting on register $B$ such that
  \begin{eqnarray}
    \| U|_T - (V \otimes W)|_T \| = \sqrt{2(1-\sqrt{p})}.
  \end{eqnarray}
\end{lemma}

\textbf{Proof:} By definition, for all states $|\psi\rangle$ in
register $A$ we have
\begin{eqnarray}
  U|\psi\rangle |\beta\rangle = \sqrt{p}V|\psi\rangle |\beta'\rangle +
  \sqrt{1-p} |\psi'\rangle |\beta''\rangle.
\end{eqnarray}
Let $W$ be a unitary operator taking $|\beta\rangle$ to
$|\beta'\rangle$.  Then
\begin{eqnarray}
  (U-V \otimes W)|\psi\rangle |\beta\rangle & = & (\sqrt{p}-1)
  V|\psi\rangle |\beta'\rangle \nonumber \\
  & & +
  \sqrt{1-p} |\psi'\rangle |\beta''\rangle.
\end{eqnarray}
Evaluating the norm, and restricting to normalized states
$|\psi\rangle |\beta\rangle$ which are in $T$, we obtain the result.

\textbf{QED}

\subsection{How noise in optical cluster-state computation maps to noise
in the quantum circuit model}
\label{subsec:ocs-threshold}

In this subsection we explain how noise in the dangling node
implementation of optical cluster-state computation is mapped to
equivalent noise in the two-at-a-time implementation described in
Subsection~\ref{subsec:ocs-intermediate}.  The results in
Subsection~\ref{subsec:ocs-intermediate} may then be used to map that
noise to equivalent noise in the original quantum circuit, which
enables us to complete the threshold proof.  Our main focus here is on
noise in single-qubit cluster-state computations, since the
multi-qubit case follows similar lines, and requires no new ideas.

\begin{figure}
  \scalebox{0.75}{\epsfig{file=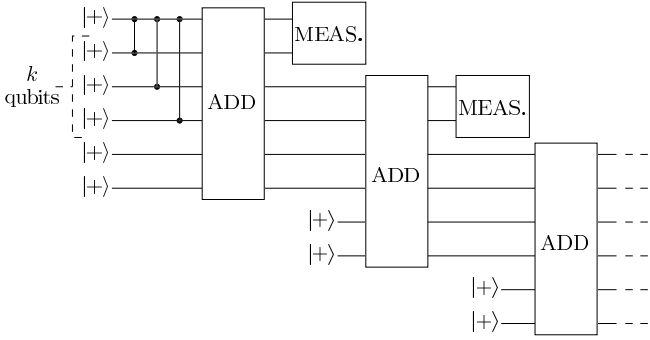}}
  \caption{The literal circuit for a dangling node implementation of a
    single-qubit cluster-state computation.  We have chosen to use an
    implementation with $k = 3$ dangling nodes; the circuit for other
    values of $k$ follows similar lines.
 \label{fig:circuit-rep-ocs}}
\end{figure}

%
% broad overview of a single-qubit cluster-state computation
%
Consider a dangling node implementation of a single-qubit
cluster-state computation.  The literal circuit for such a computation
is depicted in Figure~\ref{fig:circuit-rep-ocs}.  The first step of the
circuit sets up the initial microcluster.  The remaining steps of the
circuit alternate between applications of the operation {\sc add},
which effectively adds two extra layers of qubits to the cluster, and
the operation {\sc measure}, which implements the desired single-qubit
measurements and feedforward of measurement results.  (Note that we
have omitted ancillas, classical processing of data, and feedforward
from the visual depiction, for simplicity.)

%
% defining the add operation
%
The operation {\sc add} may be broken down into the following
operations:
\begin{itemize}
\item Preparation of microcluster states which we attempt to adjoin to
  the existing cluster.

\item All quantum gates applied in the process of attempting to adjoin
  the additional microclusters, including non-deterministic {\sc
    cphase} gates.

\item Ancillas used in the non-deterministic {\sc cphase} gates.

\item All the classical control and feedforward.

\item The computational basis measurements and local $Z$ operations
  needed to remove undesired qubits from the cluster.

\end{itemize}

\begin{figure}[hb]
\scalebox{0.85}{\epsfig{file=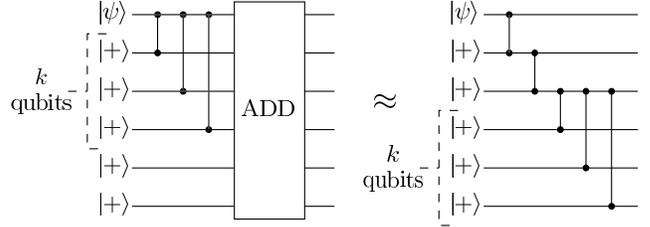}}
\caption{The operation {\sc add} satisfies the approximate circuit
identity
  illustrated here.  This identity becomes more exact as $k$ is
  increased, and the level of noise in the physical operations used to
  do the operation is decreased.
 \label{fig:approximate-circuit-identity}}
\end{figure}

In addition to these physical operations, to simplify the analysis it
is convenient to append some fictitious perfect controlled-{\sc swap}
operations to ensure that the effective output qubits always appear on
the same output lines.  That is, these controlled-{\sc swap}
operations effectively change the labels on the qubits to ensure that
the same qubits always act as the output, but otherwise don't affect
the output state.  The operations are chosen so that the approximate
circuit identity illustrated in
Figure~\ref{fig:approximate-circuit-identity} holds.  This identity
illustrates the fact that the result of attempting to add two extra
layers of qubits to the cluster of Figure~\ref{fig:microcluster-basic}
results, with high probability, in the cluster of
Figure~\ref{fig:larger-cluster-success}.  The identity fails to be
exact because (a) the attempt to add extra layers sometimes fails,
resulting (even in the ideal case of perfect operations) in the
cluster of Figure~\ref{fig:larger-cluster-failure}, and (b) the
physical operations used are inevitably somewhat noisy.

\begin{figure}
\scalebox{0.85}{\epsfig{file=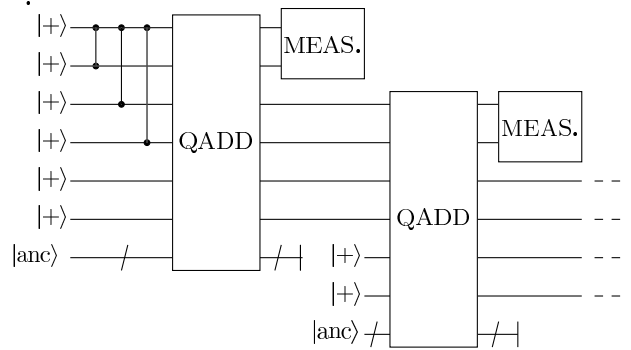}}
\caption{The output of this imperfect circuit is the same as the output
  of the noisy cluster-state computation.  The state $|{\rm
    anc}\rangle$ is an ancilla prepared (noisily) offline; it contains
  microclusters, the ancillas used in the KLM {\sc cphase}, and qubits
  used to simulate the classical control and feedforward in the
  dangling node implementation.  Note that the ancilla line is marked
  by a slash to indicate that it involves many systems, not just a
  single qubit.
 \label{fig:unitary-equivalent-ocs}}
\end{figure}

%
% make stuff unitary
%
The next step of the proof is to replace the measurements, classical
control and feedforward performed in the {\sc add} operation by
equivalent ancilla preparations and unitary quantum operations, in a
similar fashion to our threshold proof for deterministic {\sc cphase}
gates in Section~\ref{sec:deterministic}.  The result is an operation
{\sc qadd}, involving only quantum systems, perfect ancilla
preparation, and noisy unitary operations.  Note that, once again, we
model noisy ancilla preparation by perfect ancilla preparation,
followed by a noisy quantum memory step.  The output of the noisy
dangling node cluster-state computation is thus the same as the output
of the imperfect circuit illustrated in
Figure~\ref{fig:unitary-equivalent-ocs}.

\begin{figure}[hb]
\scalebox{0.75}{\epsfig{file=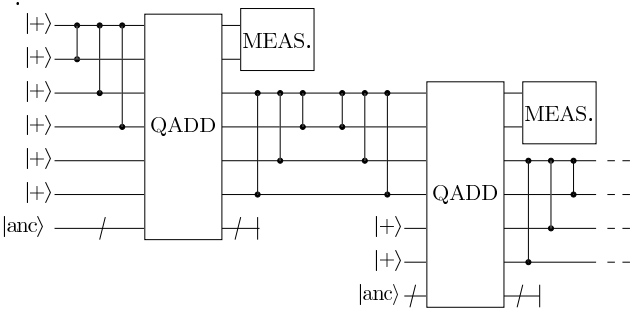}}
\caption{We may insert some fictitious perfect gates between the
operations in
  Figure~\ref{fig:unitary-equivalent-ocs}, without changing the output.
 \label{fig:ocs-with-inserted-gates}}
\end{figure}

\begin{figure}
  \scalebox{0.8}{\epsfig{file=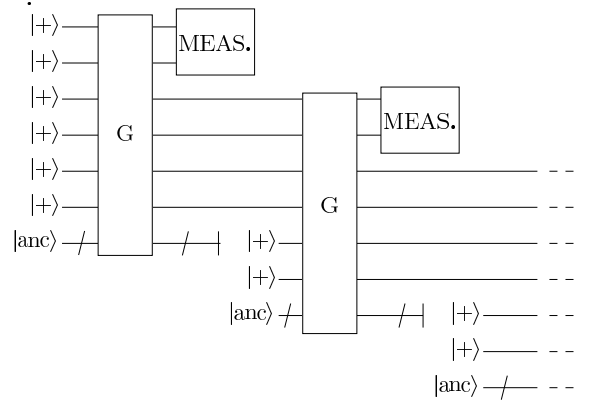}}
  \caption{The output of this imperfect circuit is the same as the
    output of the noisy dangling node cluster-state computation.
 \label{fig:repeating-circuit}}
\end{figure}

\begin{figure}[hb]
  \scalebox{0.9}{\epsfig{file=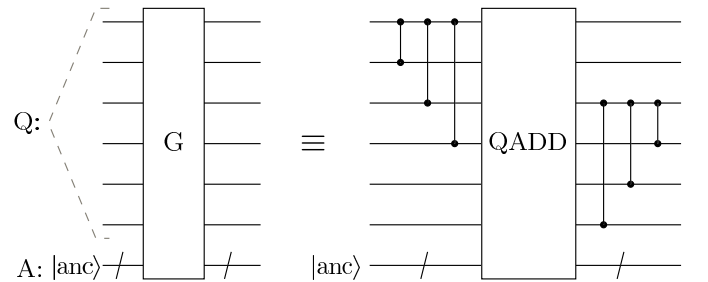}} \caption{The
    definition of the gate $G$ used in
    Figure~\ref{fig:repeating-circuit}.  This figure also defines
    subsystem labels $Q$ and $A$.
 \label{fig:define-G}}
\end{figure}

%
% add in the extra gates
%
As in the deterministic case, we may insert fictitious perfect extra
gates between {\sc qadd} operations without changing the overall
output, as illustrated in Fig~\ref{fig:ocs-with-inserted-gates}. The
entire noisy cluster-state computation is thus equivalent to the
repeating circuit illustrated in Figure~\ref{fig:repeating-circuit},
where the gate $G$ is as defined in Figure~\ref{fig:define-G}.

In our analysis it is convenient to distinguish between the perfect
gate $G$, and the imperfect gate $G_n$, which also includes the
effects of noise due to interactions with the environment, $E$.  Note
that due to the chaining property, $\Delta_{Q:A E}(G,G_n) \leq N
\eta$, where the systems $Q$ and $A$ are as defined in
Figure~\ref{fig:define-G}, $N$ is the total number of noisy operations
involved in the operation $G_n$, and $\eta$ is the maximal noise
strength of any of those operations.  Simple counting shows that $N
\leq c_1 k^2$ for some positive constant $c_1$ in the range
$10^0$-$10^2$, and so $\Delta_{Q:AE}(G,G_n) \leq c_1 k^2 \eta$.

\begin{figure}
\epsfig{file=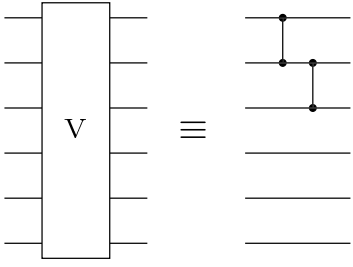} \caption{The definition of the gate
$V$.
 \label{fig:define-V}}
\end{figure}

%
% connection to postselected gates
%
With these definitions we see that $G$ is a unitary postselected gate
acting on the registers $Q$ and $A$, implementing the gate $V$ defined
in Figure~\ref{fig:define-V} on register $Q$, with probability $p_s =
1-p_f^{k-1}$.  Let $S$ be the subspace of states for $Q$ of the form
$|\psi\rangle|+\rangle \ldots |+\rangle$, where $|\psi\rangle$ is an
arbitrary single-qubit state.  Let $T$ be the subspace of states for
the system $QA$ which are of the form $|\phi\rangle |{\rm
  anc}\rangle$, where $|\phi\rangle$ is any state in $S$, and $|{\rm
  anc}\rangle$ is the initial state of the ancilla $A$.  It is also
useful to define $T'$ to be the subspace of the combined system $QAE$
containing states of the form $|\phi\rangle_{SE} \otimes |{\rm
  anc}\rangle$, where $|\phi\rangle_{SE}$ is an arbitrary state in $S
\otimes E$.  Applying Lemma~\ref{lemma:unitary-approximation}, we
conclude that there exists a unitary $W$ acting on $A$ such that
\begin{eqnarray} \label{eq:G-approx}
  \| G|_T - (V \otimes W)|_T \| = \sqrt{2(1-\sqrt{p_s})}.
\end{eqnarray}
Recalling that $\Delta_{Q:AE}(G,G_n) \leq c_1k^2 \eta$, we see that there
exists unitary $G_E$ acting on $E$ such that
\begin{eqnarray}
  \| G_n - G \otimes G_E \| \leq c_1 k^2 \eta.
\end{eqnarray}
Restricting to the subspace $T'$ we have
\begin{eqnarray} \label{eq:approx-inter-1}
  \| G_n|_{T'} - (G \otimes G_E)|_{T'} \| \leq c_1 k^2 \eta.
\end{eqnarray}
{}From Eq.~(\ref{eq:G-approx}) we deduce that
\begin{eqnarray}
  \| (G \otimes G_E)|_{T'} - (V \otimes W \otimes G_E)|_{T'} \|
  = \sqrt{2(1-\sqrt{p_s})}. \nonumber \\ \label{eq:approx-inter-2}
& &
\end{eqnarray}
Using the triangle inequality and Eqs.~(\ref{eq:approx-inter-1})
and~(\ref{eq:approx-inter-2}) we obtain
\begin{eqnarray}
  \| G_n|_{T'} - (V \otimes W \otimes G_E)|_{T'} \| \leq c_1 k^2 \eta
  + \sqrt{2(1-\sqrt{p_s})}. \nonumber \\
 & &
\end{eqnarray}
Applying the second unitary extension theorem, Theorem~\ref{thm:uet2},
we see that there exists unitary $\tilde G_n$ whose action on $T'$ is
identical to the action of $G_n$, and such that
\begin{eqnarray}
  \| \tilde G_n - (V \otimes W \otimes G_E) \| & \leq & 2c_1 k^2 \eta
  + 2\sqrt{2(1-\sqrt{p_s})}. \nonumber \\
 & &
\end{eqnarray}

\begin{figure}
  \epsfig{file=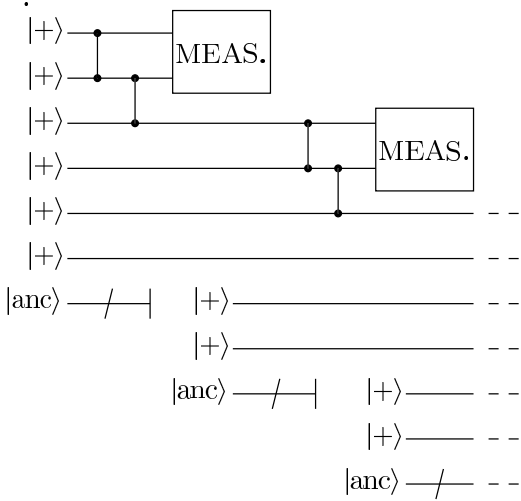}
  \caption{The output of this noisy circuit is the same as the output
    of the noisy cluster-state computation.
 \label{fig:equivalent-repeating-circuit}}
\end{figure}

Summarizing, the output of the noisy cluster-state computation is the
same as the output of the noisy circuit in
Figure~\ref{fig:equivalent-repeating-circuit}.  The blocks of two
imperfect {\sc cphase} gates shown in this circuit represent the
operation $\tilde G_n$, which we have seen satisfies
\begin{eqnarray}
  \Delta_{Q:A E}(V,\tilde G_n) \leq 2c_1 k^2 \eta
  + 2\sqrt{2(1-\sqrt{p_s})},
\end{eqnarray}
where $p_s = 1-p_f^{k-1}$ is the probability of successfully adjoining
a microcluster, and $p_f$ is the probability of a non-deterministic
{\sc cphase} failing.  Examining
Figure~\ref{fig:equivalent-repeating-circuit}, we see that a dangling
node implementation using non-deterministic gates is equivalent to a
noisy two-at-a-time deterministic implementation, where the additional
layers of the cluster are added with noise of total strength at most
$2c_1 k^2 \eta + 2\sqrt{2(1-\sqrt{p_s})}$.  Combining this with the
results about the two-at-a-time implementation in
Subsection~\ref{subsec:ocs-intermediate}, we see that a noisy dangling
node implementation of a single-qubit cluster-state computation is
equivalent to a noisy single-qubit quantum circuit computation.  Noise
of strength $\eta$ in the dangling node implementation is mapped to
equivalent noise of strength $c_1 k^2 \eta +c_2 \eta +
2\sqrt{2(1-\sqrt{p_s})}$ in the quantum circuit model of computation,
where the extra contribution $c_2 \eta$ is due to noise in the
single-qubit measurements, and $c_2$ is a positive constant of order
$10^0$.

%
% multi-qubit case
%
Similar reasoning can be used to map noise of strength $\eta$ in a
dangling node implementation of a multi-qubit cluster-state
computation back to the original quantum circuit with noise of
strength at most $c_1 k^2 \eta + c_2 \eta + 2\sqrt{2(1-\sqrt{p_s})}$,
where the constants $c_1$ and $c_2$ may now be different, but are
still of the same order, and $p_s = 1-2p_f^{k-1}+p_f^{2k-2}$.  The
only difference in the proof is that in addition to the $G$
operations, multi-qubit computations also involve analogous operations
based on the process of adjoining the more complex microcluster of
Figure~\ref{fig:other-microcluster}.  However, the analysis for such
operations goes through in exactly the same way as in the single-qubit
case.

%
% summing up
%
Summing up, we have shown that noise of strength up to $\eta$ in a
dangling node implementation of a cluster-state computation is
equivalent to noise of strength at most $c_1 k^2 \eta + c_2 \eta + 2
\sqrt{2(1-\sqrt{p_s})}$ in the original quantum circuit, where $c_1$
is a positive constant of order $10^0$-$10^2$, $c_2$ is a positive
constant of order $10^0$, and $p_s = 1-2p_f^{k-1}+p_f^{2k-2}$.  Thus,
provided $k$ and $\eta$ satisfy
\begin{eqnarray}
   c_1 k^2 \eta + c_2 \eta + 2\sqrt{2(1-\sqrt{p_s})} \leq \eta_{\rm th}
\end{eqnarray}
it is possible to compute fault-tolerantly in the optical
cluster-state proposal for computation.  This may be rephrased as the
condition:
\begin{eqnarray}
   \eta \leq \eta_{k,{\rm th}}^{\rm ocs} \equiv
   \frac{\eta_{\rm th}-2 \sqrt{2(1-\sqrt{p_s})}}{c_1 k^2+c_2}.
\end{eqnarray}
We can always ensure that $\eta_{k,{\rm th}}^{\rm ocs} > 0$ by
choosing $k$ sufficiently large.  Provided this condition is
satisfied, $\eta_{k,{\rm th}}^{\rm ocs}$ is thus a threshold for
optical cluster-state computation.

%% file: ftqc-cs-conclusion.tex
\section{Conclusion}
\label{sec:conclusion}

In this paper we have proved two fault-tolerant threshold theorems for
the cluster-state model of quantum computation.  Our first threshold
theorem applies to implementations in which deterministic (but noisy)
entangling gates are available.  Our second threshold theorem is
specifically adapted to the case of optical quantum computation, where
entangling gates are performed non-deterministically.  In both cases
our threshold theorems hold for quite pessimistic noise models,
allowing non-Markovian noise applied by an intelligent adversary who
can exploit constructive interference to enhance the effects of the
noise, and even cause errors in the classical computation and
feedforward of measurement results.  A drawback of our noise models is
that they do not yet allow the possibility of leakage errors, like
photon loss in optics.  We expect to remove this drawback in future
work, by combining the ideas in~\cite{Terhal04a} with well-established
methods for dealing with photon loss, e.g.~\cite{Knill00a}.

Our focus has been on proving that a finite threshold \emph{exists}
for cluster-state computation, rather than on obtaining a precise
numerical evaluation of the threshold.  This is consistent with our
general philosophy of understanding the threshold through a two-part
process: first, rigorously proving the existence of a finite threshold
for some large class of noise models; and second, through a
combination of numerical and analytic work obtaining a realistic
estimate of the threshold for some specific and physically-motivated
noise model. In this paper we have obtained a rigorous proof that a
finite threshold exists.  Detailed numerical simulation and
optimization of the threshold value for realistic noise models is
underway, and will be reported elsewhere~\cite{Dawson04b}.

Our investigations in this paper have been geared toward variants of
the one-way quantum computing model introduced by Raussendorf and
Briegel~\cite{Raussendorf01a}.  However, the techniques we have
proposed seem quite generally applicable to the task of making
measurement-based schemes for quantum computation fault-tolerant.  It
seems likely that schemes such as those proposed
in~\cite{Nielsen03b,Yoran03a,Jorrand04a,Childs04a} (see also
references therein) can be made fault-tolerant using similar ideas.
Particularly appealing from a theoretical point of view is the
possibility of fault-tolerant computation using measurement alone,
with no unitary gates whatsoever (excepting quantum memory).  This
might be done using, for example, a scheme such as~\cite{Nielsen03b},
or one of the simpler variants that has since been
proposed~\cite{Fenner01a,Leung01c,Leung03a,Perdrix04a}.

The most important conclusion from our results is that noise need not
be an obstacle to scalable quantum computation using cluster states.
In particular, our results provide encouraging evidence that practical
 proposals for cluster-state quantum computation using neutral
atoms~\cite{Raussendorf01a} and optics~\cite{Nielsen04b} are, in
principle, fully scalable approaches to quantum computation.  

\acknowledgments

Thanks to Andrew Childs, Jennifer Dodd, Andrew Doherty, Alexei
Gilchrist, Henry Haselgrove, Debbie Leung, Gerard Milburn, Tim Ralph,
Rob Spekkens and Andrew White for enjoyable and informative
discussions.  Thanks in particular to Alexei Gilchrist for stressing
to us the importance of leakage errors associated with photon loss in
optics.  Thanks also to Hans Briegel and Robert Raussendorf for
correspondence relating to their work on fault-tolerance and
measurement-based quantum computing.

%% file: ftqc-cs-app1.tex
\section{The unitary extension theorems}
\label{app:unitary-extension}

%
% what we do
%
In this appendix we restate and prove the first and second unitary
extension theorems, from Section~\ref{subsec:error-strength}.

\begin{theorem} {} \textbf{(First unitary extension theorem)}
  Let $U, \tilde U$ and $V$ be unitaries acting on a Hilbert space
  $T$.  Suppose $S$ is a subspace of $T$ such that $U$ and $\tilde U$
  have the same action on $S$, i.e., $U|_S = \tilde U|_S$.  (Note that
  we do not assume that $U$ and $\tilde U$ leave the subspace $S$
  invariant, so $U|_S$ and $\tilde U|_S$ should be considered as maps
  from $S$ into $T$.)  Then there exists a unitary extension $\tilde
  V$ of $V|_S$ to the entire space $T$ such that
  \begin{eqnarray}
    \| \tilde V - \tilde U \| \leq \| V-U \|.
  \end{eqnarray}
\end{theorem}

It is worth noting that the proof below holds not just for the matrix
norm, but for any norm such that $\| A B \| \leq \| A \| \, \| B \|$, and
$\| W \| \leq 1$ for all unitaries $W$.

\textbf{Proof:} Let $P$ be the projector onto the subspace $S$, and
let $Q = I-P$ be the projector onto the orthocomplement of $S$ in $T$.
Define
\begin{eqnarray} \label{eq:extension-definition}
  \tilde V \equiv VP + VU^\dagger \tilde U Q.
\end{eqnarray}
We will show that $\tilde V$ has the required properties.  It is clear
that $\tilde V$ is an extension of $V|_S$.  To prove unitarity of
$\tilde V$ we first observe that
\begin{eqnarray} \label{eq:unitary-critical}
  U^\dagger \tilde U Q = Q U^\dagger \tilde U.
\end{eqnarray}
To prove this equation, observe that it is equivalent to $U^\dagger
\tilde U P = P U^\dagger \tilde U$, since $Q = I-P$.  But $U^\dagger
\tilde U P = P U^\dagger \tilde U$ follows easily from the fact that
$U|_S = \tilde U|_S$.  The unitarity of $\tilde V$ follows from
Eqs.~(\ref{eq:extension-definition}),~(\ref{eq:unitary-critical}), and
some algebra:
\begin{eqnarray}
  \tilde V \tilde V^\dagger & = & VP V^\dagger + V (U^\dagger \tilde U) Q
  (U^\dagger \tilde U)^\dagger V^\dagger \\
  & = & VPV^\dagger + VQV^\dagger \\
  & = & VV^\dagger \\
  & = & I.
\end{eqnarray}
To bound $\| \tilde V - \tilde U \|$ observe that
\begin{eqnarray}
  \tilde V - \tilde U & = & VP+VU^\dagger \tilde U Q - \tilde U P - \tilde U Q.
\end{eqnarray}
Observing that $\tilde U P = UP$ and inserting $UU^\dagger = I$ we get
\begin{eqnarray}
  \tilde V - \tilde U & = & VP+VU^\dagger \tilde U Q -UP - UU^\dagger
  \tilde U Q \\
  & = & (V-U)(P+U^\dagger \tilde U Q).
\end{eqnarray}
Recall that $\| A B \| \leq \| A \| \, \| B \|$, and observe that
$P+U^\dagger \tilde U Q$ is unitary.  It follows that $\| P +
U^\dagger \tilde U Q \| = 1$, and thus $\| \tilde V - \tilde U \| \leq
\| V- U\|$, as required.

\textbf{QED}

\begin{theorem} {} \textbf{(Second unitary extension theorem)}
  Let $U$ and $V$ be unitary operations acting on a
  (finite-dimensional) inner product space $T$.  Suppose $S$ is a
  subspace of $T$.  Then there exists a unitary operation $\tilde V$
  such that $\tilde V|_S = V|_S$ and
  \begin{eqnarray}
    \| U - \tilde V\| \leq 2 \| U|_S - V|_S\|.
  \end{eqnarray}
\end{theorem}

%
% how we prove this
%
The proof of the second unitary extension theorem is somewhat more
complex than the proof of the first.  We begin the proof by
introducing some notation related to the singular value decomposition
of a matrix, and then we state and prove some simple lemmas about
singular values and matrix norms.

%
% the singular value decomposition
%
Recall that the singular value decomposition states that an arbitrary
$m \times n$ matrix $M$ can be written $M = L_M \Sigma_M R_M$, where
$L_M$ is an $m \times m$ unitary matrix, $R_M$ is an $n \times n$
unitary matrix, and $\Sigma_M$ is an $m \times n$ matrix, all of whose
entries are zero, except the diagonal entries $(\Sigma_M)_{jj} =
\sigma_j(M)$, known as the \emph{singular values}, which are
non-negative and arranged in decreasing order, $\sigma_1(M) \geq
\sigma_2(M) \geq \ldots$.  It is easy to see that the singular values
are determined by the equation $\sigma_j(M)^2 = \lambda_j(M^\dagger M)
= \lambda_j(M M^\dagger)$, the $j$th largest eigenvalues of the
matrices $M^\dagger M$ and $M M^\dagger$.  It is also useful to note
that $\| M \| = \sigma_1(M)$. When $M$ is an $m \times m$ matrix
(i.e., a square matrix), we will use the notation $\sigma_{\min}(M)
\equiv \sigma_m(M)$ to denote the smallest singular value of $M$.

\begin{proposition} Suppose $U$ is a unitary matrix that can be written
  \begin{eqnarray}
    U = \left[ \begin{array}{cc} A & C \\ B & D \end{array} \right],
  \end{eqnarray}
  where $A$ is an $m \times m$ matrix, $B$ is $n \times m$, $C$ is $m
  \times n$, and $D$ is $n \times n$.  Then $\sigma_{\min}(A) =
  \sigma_{\min}(D)$.
\end{proposition}

\textbf{Proof:} Inspection of the equations $U^\dagger U = U U^\dagger
= I$ implies that $A^\dagger A + B^\dagger B = I_m$ and $BB^\dagger +
DD^\dagger = I_n$, where we use $I_k$ to denote the $k \times k$
identity matrix.  Using these equations we have:
\begin{eqnarray}
  \sigma_{\min}(A)^2 & = & \lambda_m(A^\dagger A) \\
  & = & 1-\lambda_1(B^\dagger B) \\
  & = & 1-\lambda_1(B B^\dagger) \\
  & = & \lambda_n(DD^\dagger) \\
  & = & \sigma_{\min}(D)^2,
\end{eqnarray}
from which the result follows.  An alternate proof of this proposition
follows immediately from the well-known CS decomposition of linear
algebra (see, for example, Theorem~VII.1.6 on page~196
of~\cite{Bhatia97a}).  

\textbf{QED}

To state the next proposition we need to introduce some additional
notation.  We define the partial order $X \leq Y$ for matrices $X$ and
$Y$ if $Y-X$ is a positive matrix.  We define $|X| \equiv
\sqrt{X^\dagger X}$.

\begin{proposition} {} 
  Suppose $M$ is a matrix such that $|M| \leq I$.  Then $\| I- M\|
  \geq 1-\sigma_{\min}(M)$.
\end{proposition}

\textbf{Proof:} Choose a normalized vector $|\psi\rangle$ so that
$\Sigma_M |\psi\rangle = \sigma_{\min}(M)|\psi\rangle$.  By the singular
value decomposition $\|I - M\| = \|I-L_M \Sigma_M R_M \| = \|
W-\Sigma_M\|$, where $W \equiv L_M^\dagger R_M^\dagger$.  Thus
\begin{eqnarray}
  \|I-M\| & \geq & \| (W-\Sigma_M)|\psi\rangle \| \\
  & \geq & \| W|\psi\rangle \| -\| \Sigma_M|\psi\rangle \| \\
  & = & 1-\sigma_{\min}(M),
\end{eqnarray}
where we have used the triangle inequality.

\textbf{QED}

\begin{proposition} {} \label{prop:Aronszajn}
  Let
  \begin{eqnarray}
    M = \left[ \begin{array}{cc} A & B \\ B^\dagger & C \end{array}
      \right]
  \end{eqnarray}
  be a positive (and thus Hermitian) square matrix.  We assume $A$ is
  $m \times m$, $B$ is $m \times n$, and $C$ is $n \times n$.  Then
  \begin{eqnarray}
    \| M \| \leq \| A \|+\| C\|.
  \end{eqnarray}
\end{proposition}

\textbf{Proof:} This proposition can be viewed as a special case of
Aronszajn's inequality, Theorem III.2.9 on page~64
of~\cite{Bhatia97a}.  Alternately, since $M$ is positive we can find a
block matrix $D = [D_1 D_2]$ such that $M = D^\dagger D$, and thus $A
= D_1^\dagger D_1$ and $C = D_2^\dagger D_2$.  We have
\begin{eqnarray}
  \| M \| & = & \lambda_1(D^\dagger D) \\
  & = & \lambda_1(D D^\dagger) \\
  & = & \lambda_1(D_1 D_1^\dagger+D_2 D_2^\dagger) \\
  & \leq & \lambda_1(D_1 D_1^\dagger) + \lambda_1(D_2 D_2^\dagger) \\
  & = & \| A\| + \| B\|,
\end{eqnarray}
where the second-last line follows from the well-known eigenvalue
inequality $\lambda_1(A+B) \leq \lambda_1(A)+\lambda_1(B)$, true for
all Hermitian matrices $A$ and $B$.

\textbf{QED}

\textbf{Proof of the second unitary extension theorem:} We prove the
theorem in two parts.  In the first part we prove the result for the
case $U = I$.  In the second part we show that the general result
follows from the case when $U = I$.

For the first part, we write $V$ in block-diagonal form as
\begin{eqnarray}
  V = 
  \left[ \begin{array}{cc} A & C \\ B & D \end{array}
    \right],
\end{eqnarray}
where the first block represents a basis for $S$, and the second block
represents a basis for the orthocomplement $S_\perp$.  Let the
singular value decomposition of $D$ be $D = L_D \Sigma_D R_D$.
Rotating by $L_D^\dagger$ to change the basis of $S_\perp$ we see that
$V$ can be written in the new basis as
\begin{eqnarray}
  V = \left[ \begin{array}{cc} A' & C' \\ B' & \Sigma_D R_D L_D
      \end{array} \right],
\end{eqnarray}
where $A', B',C'$ represent the action of $V$ with respect to the new
basis.  We define the extension $\tilde V$ by
\begin{eqnarray}
  \tilde V & \equiv & V \left[ \begin{array}{cc} I & 0 \\
      0 & L_D^\dagger R_D^\dagger \end{array} \right] \\
 & = & \left[ \begin{array}{cc} A' & C'' \\ B' & \Sigma_D
     \end{array} \right],
\end{eqnarray}
where $C'' = C' L_D^\dagger R_D^\dagger$.  It is clear that $\tilde
V$ is unitary and $\tilde V|_S = V_S$. All that remains to complete
the first part of the proof is to bound $\|\tilde V - I \|$.  We have
\begin{eqnarray}
  \| \tilde V - I \|^2 & = & \| (\tilde V-I)(\tilde V^\dagger - I) \| \\
 & = & \| 2I -\tilde V-\tilde V^\dagger \|  \\
 & = & \left\| \left[ \begin{array}{cc} 2I-A'-A'^\dagger &
    -C''-B'^\dagger \\
    -B'-C''^\dagger & 2I-2\Sigma_D \end{array} \right] \right\|. \nonumber \\
 & &
\end{eqnarray}
Applying Proposition~\ref{prop:Aronszajn} we obtain
\begin{eqnarray}
  \| \tilde V-I\|^2 \leq \| 2I-A'-A'^\dagger\|
    + \| 2I-2\Sigma_D \|.
\end{eqnarray}
But $\| 2I-A'-A'^\dagger\| \leq \|I-A'\| + \|I-A'^\dagger \| =
2\|I-A'\| \leq 2\|I|_S - V|_S\|$.  We also have $\| 2I-2\Sigma_D\| =
2-2\sigma_{\min}(D) = 2-2\sigma_{\min}(A') \leq 2 \|I-A'\| \leq 2
\|I|_S-V|_S\|$.  The first part of the proof follows.

For the second part of the proof we set $U' \equiv I, V' \equiv
U^\dagger V$.  Then by the first part of the proof there exists
unitary $\tilde V'$ such that $\tilde V'|_S = U^\dagger V|_S$, and $\|
\tilde V'-I\| \leq 2 \| U^\dagger V|_S-I_S\|$.  Set $\tilde V \equiv
U\tilde V'$.  Then $\tilde V$ is a unitary matrix such that $\tilde
V|_S = V|_S$ and
\begin{eqnarray}
  \| \tilde V-U \| & = & \| \tilde V'-I \| \\
  & \leq & 2 \| U^\dagger V|_S-I|_S\| \\
  & = & 2 \| V|_S-U|_S\|.
\end{eqnarray}

\textbf{QED}

An alternate proof of the second unitary extension theorem may be
given, following similar lines, but based on the well-known CS
decomposition from linear algebra.  We have taken the approach
presented here as it is only slightly more complex than the alternate
proof, and relies on less background material.

%% file: ftqc-cs-app2.tex
\section{Proof of Proposition~\ref{prop:quantum-circuit-identity}}
\label{app:single-qubit-identity}

\newtheorem{identity}{Identity}

In this appendix we prove
Proposition~\ref{prop:quantum-circuit-identity}, which asserts the
truth of the circuit identity in Figure~\ref{fig:qb-equiv-hz}.  This
proposition establishes the correspondence between blocks $Q B_\alpha$
and gates $H Z_\alpha$ in the quantum circuit.

To prove Proposition~\ref{prop:quantum-circuit-identity} we make use
of four simple circuit identities which we assert here without proof.
Each may be readily verified by direct computation. The first identity
is a simplification of a common element in literal circuits for
cluster-state computation.

% --------------------------------------------------------------------- %
% IDENTITY: Shorthand for U_\alpha with CNOT on rhs                     %
\begin{identity}
The following circuit simplification holds where the second qubit is in
the state $|+\rangle$.
\begin{center}
\epsfig{file=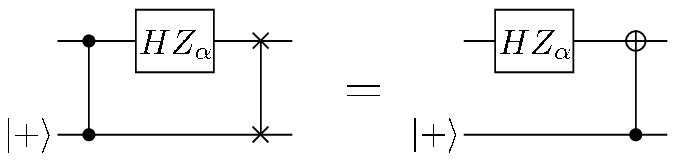}
\end{center}
\label{identity:A}
\end{identity}
% --------------------------------------------------------------------- %

The remaining three identities concern the commutativity properties of
{\sc cnot} (controlled-{\sc not}) and {\sc cphase} gates when the
control of one operator is the target of the operator immediately
following or preceding it.

% --------------------------------------------------------------------- %
% IDENTITY: Commuting overlapping CNOTS                                 %
\begin{identity}
The following circuit identity holds:
\begin{center}
\epsfig{file=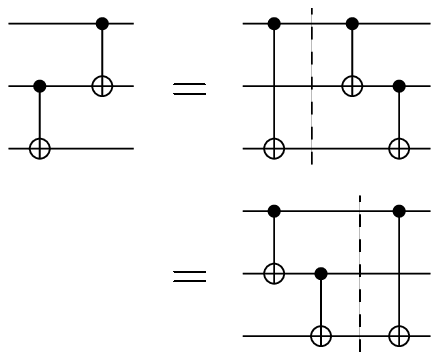}
\end{center}
\label{identity:B}
\end{identity}
% --------------------------------------------------------------------- %

% --------------------------------------------------------------------- %
% IDENTITY: Commuting overlapping CNOT and CPHASE                       %
\begin{identity}
The following circuit identity holds:
\begin{center}
\epsfig{file=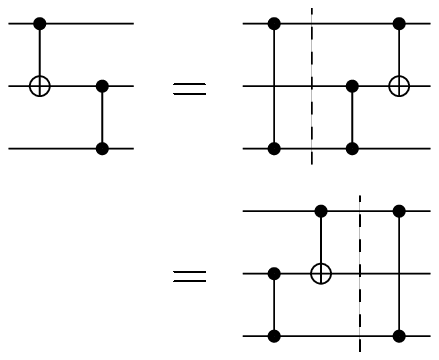}
\end{center}
\label{identity:C}
\end{identity}
% --------------------------------------------------------------------- %

% --------------------------------------------------------------------- %
% IDENTITY: Other ordering of the previous identity                       %
\begin{identity}
The following circuit identity holds:
\begin{center}
\epsfig{file=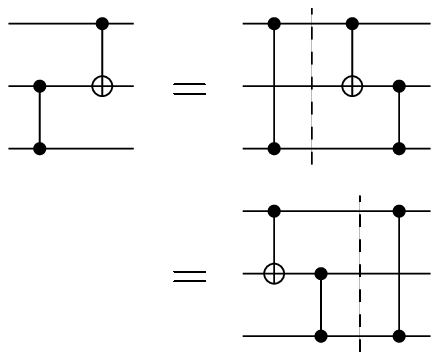}
\end{center}
\label{identity:D}
\end{identity}
% --------------------------------------------------------------------- %

\begin{figure*}
\epsfig{file=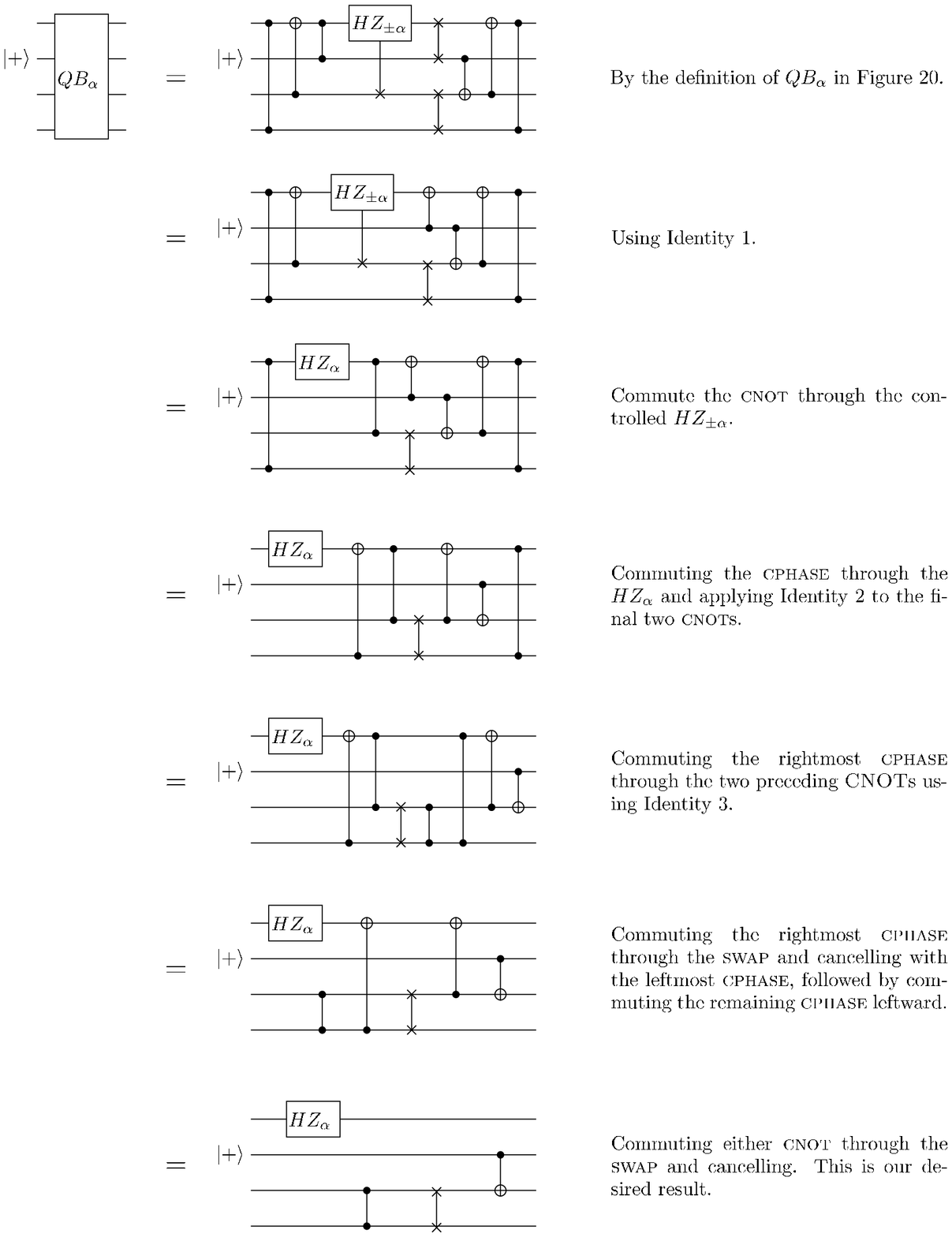} \caption{The sequence of
circuit identities used to prove
Proposition~\ref{prop:quantum-circuit-identity}
\label{fig:circuit-proof}}
\end{figure*}

With these identities we can now prove
Proposition~\ref{prop:quantum-circuit-identity}.  The proof is a short
sequence of circuit identities which are shown in
Figure~\ref{fig:circuit-proof}.

%\begin{proposition} {}  \label{prop:quantum-circuit-identity-repeat}
%The circuit identity of Figure~\ref{fig:qb-equiv-hz} holds, where both
%circuits are assumed to be perfect.  All inputs are assumed to be
%arbitrary, except the fixed $|+\rangle$ input, as shown.
%\end{proposition}

% --------------------------------------------------------------------- %
% FIGURE: Circuit diagram proof of prop:quantum-circuit-identity        %
% --------------------------------------------------------------------- %